\documentclass[10pt,journal,compsoc]{IEEEtran}

\ifCLASSOPTIONcompsoc
  \usepackage[nocompress]{cite}
\else
  \usepackage{cite}
\fi

\usepackage{soul}
\usepackage{amsmath}
\usepackage{amsthm}
\usepackage{amssymb}
\usepackage{amsfonts}
\usepackage{textcomp}
\usepackage{xcolor}
\usepackage{array,float,url,graphicx,wrapfig,verbatim,color,comment,caption}
\usepackage{subcaption}
\usepackage{amsfonts,bm}
\usepackage{algorithm,algorithmic}
\usepackage{amssymb}
\usepackage{pgfgantt}
\usepackage{multirow}
\usepackage{soul}
\usepackage{bbding}
\usepackage{enumerate}
\usepackage{booktabs}
\usepackage{tabularx}
\usepackage{bbm}
\usepackage{xr}
\usepackage{siunitx}
\usepackage{adjustbox}
\makeatletter
\DeclareMathOperator{\topk}{top}

\newtheorem{lemma}{\textbf{Lemma}}
\newtheorem{theorem}{\textbf{Theorem}}

\newtheorem{assumption}{\textbf{Assumption}}
\theoremstyle{definition}
\newtheorem{remark}{\textbf{Remark}}

\newcounter{relctr} %% <- counter for relations
\everydisplay\expandafter{\the\everydisplay\setcounter{relctr}{0}} %% <- reset every eq
\newcommand\labelrel[2]{%
  \begingroup
    \refstepcounter{relctr}%
    \stackrel{\textnormal{(\alph{relctr})}}{\mathstrut{#1}}%
    \originallabel{#2}%
  \endgroup
}
\AtBeginDocument{\let\originallabel\label} %% <- store original definition

\ifCLASSINFOpdf
\else
\fi

\begin{document}
\title{Heterogeneity-Aware Cooperative Federated Edge Learning with Adaptive Computation and Communication Compression}
\author{Zhenxiao~Zhang,~\IEEEmembership{Student~Member,~IEEE,} Zhidong~Gao,~\IEEEmembership{Student~Member,~IEEE,} Yuanxiong~Guo,~\IEEEmembership{Senior~Member,~IEEE,} and Yanmin~Gong,~\IEEEmembership{Senior~Member,~IEEE}
\IEEEcompsocitemizethanks{\IEEEcompsocthanksitem Z. Zhang, Z. Gao, and Y. Gong are with the Department of Electrical and Computer Engineering, The University of Texas at San Antonio, San Antonio, TX, 78249. Y. Guo is with the Department of Information Systems and Cyber Security, The University of Texas at San Antonio, San Antonio, TX, 78249. E-mail: \{zhenxiao.zhang@my., zhidong.gao@my., yuanxiong.guo@, yanmin.gong@\}utsa.edu.}
% \IEEEcompsocitemizethanks{\IEEEcompsocthanksitem Z. Zhang and Z. Gao both contributed equally to this work.}
}

\markboth{}%
{Zhang \MakeLowercase{\textit{et al.}}: Heterogeneity-Aware Cooperative Federated Edge Learning with Adaptive Computation and Communication Compression}

\IEEEtitleabstractindextext{%
\begin{abstract}
Motivated by the drawbacks of cloud-based federated learning (FL), cooperative federated edge learning (CFEL) has been proposed to improve efficiency for FL over mobile edge networks, where multiple edge servers collaboratively coordinate the distributed model training across a large number of edge devices. However, CFEL faces critical challenges arising from dynamic and heterogeneous device properties, which slow down the convergence and increase resource consumption. This paper proposes a heterogeneity-aware CFEL scheme called \textit{Heterogeneity-Aware Cooperative Edge-based Federated Averaging} (HCEF) that aims to maximize the model accuracy while minimizing the training time and energy consumption via adaptive computation and communication compression in CFEL. By theoretically analyzing how local update frequency and gradient compression affect the convergence error bound in CFEL, we develop an efficient online control algorithm for HCEF to dynamically determine local update frequencies and compression ratios for heterogeneous devices. Experimental results show that compared with prior schemes, the proposed HCEF scheme can maintain higher model accuracy while reducing training latency and improving energy efficiency simultaneously.  
\end{abstract}

}

\maketitle

\IEEEdisplaynontitleabstractindextext
\IEEEpeerreviewmaketitle

\IEEEraisesectionheading{\section{Introduction}\label{sec:intro}}
As a distributed machine learning (ML) paradigm that has gained significant attention recently, federated learning (FL) allows edge devices to collaboratively learn a shared model while keeping their training data locally, which offers several advantages such as privacy protection and better efficiency compared to centralized ML paradigm \cite{mcmahan2017communication}. When deployed over mobile edge networks \cite{zhang2024energy,yu2020joint,ding2018beef}, the standard FL framework is cloud-based and relies on a central cloud server to coordinate the learning process and aggregate model updates from all edge devices. However, due to the long-distance and limited-bandwidth transmissions between an edge device and the remote cloud, the model training in cloud-based FL is inevitably slow and fails to meet the latency requirements of delay-sensitive intelligent applications~\cite{kairouz2021advances}. Moreover, the reliance on a central server presents the risk of a single point of failure in cloud-based FL.

With the increasing deployment of computation and storage resources at the edge in 5G-and-beyond networks, an edge-based FL framework called cooperative federated edge learning (CFEL) is emerging as a promising alternative to cloud-based FL~\cite{castiglia2021multi,sun2023semi,zhong2021p,zhang2022scalable}. In this framework, edge servers co-located with the mobile base stations are responsible for the coordination among their proximate edge devices. Edge servers only communicate with their neighboring servers to avoid the communication bottleneck at the central cloud. Moreover, by eliminating the need for a single server to collect local model updates from all devices, CFEL can effectively mitigate the risk of a single point of failure and provide a scalable FL framework for vast numbers of edge devices distributed across a wide area.

Despite its immense potential, CFEL faces several challenges that hinder training efficiency: 1)~\textit{System Heterogeneity}. In the CFEL system, the edge devices are notably heterogeneous, displaying a wide range of capabilities. For example, there can be substantial differences in computing power, such as CPU frequency and battery life, with some edge devices being up to ten times more powerful than others. Similarly, the communication capabilities, including bandwidth and throughput, also vary significantly among these edge devices. 2) \textit{Dynamic State.} Since the environment of participating devices may fluctuate over time, the available resources (e.g., CPU states, channel states, and battery life) for training on edge devices could be dynamic. Due to system heterogeneity and dynamic state, the computation and communication resources may vary significantly across all devices, leading to the straggler problem during the training with large training delay and high energy consumption. How to achieve efficient model training under system heterogeneity and dynamic state in CFEL remains largely unknown.

To address the aforementioned challenges, in this paper, we propose a novel learning scheme called \textit{Heterogeneity-Aware Cooperative Edge-based Federated Averaging} (HCEF) to achieve both high-accuracy and resource-efficient model training in CFEL. Specifically, at each communication round, devices update their local models by adapting their local update frequencies to eliminate the idle time and further compress their local updates using personalized compression ratios to enhance communication efficiency. Through this joint optimization approach, HCEF can achieve a high model accuracy while minimizing the training latency and energy consumption, making it well-suited for mobile edge networks with limited resources. 

While previous research~\cite{nori2021fast,xu2022adaptive} in traditional FL has explored the joint optimization of local update frequency and model compression ratio to enhance training efficiency and reduce resource consumption, our study differs significantly in several challenging respects due to the decentralized network communication topology in CFEL. 
First,  the relationship between local update frequency, model compression ratio, model accuracy, training time, and energy consumption within CFEL has not been thoroughly investigated. This introduces a new challenge for our research - to explore the underlying connections between these factors. Second, within the CFEL framework, deploying a learning algorithm that maximizes model accuracy while minimizing both training time and energy consumption, using adaptive computation and communication compression, presents another significant challenge. 

In summary, the main contributions of this paper are as follows:
\begin{itemize}
    \item We propose an efficient learning scheme named HCEF, which incorporates adaptive control of local update frequency and communication compression to effectively address the challenges posed by system heterogeneity and dynamic state in CFEL.

    \item By theoretically analyzing how adaptive local update frequency and communication compression affect the convergence error bound, we formulate an optimization problem that jointly optimizes local update frequencies and compression ratios to achieve high accuracy subject to time and energy constraints. 
    
    \item We develop an algorithm to efficiently solve the formulated problem, which is generally non-convex, by alternatively optimizing local update frequencies and compression ratios.

    \item We evaluate our scheme through extensive experiments based on common FL benchmark datasets and demonstrate that HCEF can learn an accurate model with a shorter training time and lower energy consumption than other FL schemes at mobile edge networks. 
\end{itemize}

The rest of this paper is organized as follows. Section~\ref{sec:related_work} reviews related works. Section~\ref{sec:prel_prob} introduces the system model and formulates the optimization problem. The main convergence result is introduced in Section~\ref{sec_conv}. Section~\ref{sec:alg_design} reformulates the optimization problem and provides an efficient algorithm to solve it. Section~\ref{sec:exp} shows the experimental results, and Section~\ref{sec:conclusion} concludes the paper.

%%%%%%%%%%%%%%%%%%%%%%%%%%%%%%%%%%%%%%%%%%%%%
\section{Related works}\label{sec:related_work}
%%%%%%%%%%%%%%%%%%%%%%%%%%%%%%%%%%%%%%%%%%%%%
%----------------------------------------
\begin{table}[t]
  \caption{Summary of main notations.}
  \label{tab:notations}
  \centering
  \begin{tabular}{cc}
    \toprule
    Notation & Definition\\
    \midrule
    $i, j$ &  Index for cluster\\
    $n$ & Index for device\\
    $l$ & Index for global round\\
    $r$ & Index for edge round\\
    $s$ & Index for local iteration\\
    $t$ & Index for global iteration\\
    $N$ & Total number of devices\\
    $m$ & Total number of edge servers/clusters\\
    % $[m]$ & \{1, 2, \ldots, $m$\}\\
    $\mathcal{S}_i$ & Set of devices in cluster $i$\\
    $\mathcal{S}$ & Set of all devices\\
    $N_i$ & Number of devices in cluster $i$\\
    $\mathcal{G}$ & Communication graph for edge backhaul\\ 
    $\mathbf{y}_i^{l, r}$ &  Edge model of cluster $i$\\ 
    $\mathcal{D}_n$ & Data distribution of device $n$\\
%    $z$ & A data sample\\
    $F_n(\cdot)$ & Local objective function of device $n$\\
    $\rho_n$ & Update probability of device $n$ \\
    $\theta_n$ & Compression ratio of device $n$ \\
    $\mathbf{x}_n^{l, r, s}$ & Local model of device $n$\\
    $\mathbf{g}_n$ & Stochastic gradient of device $n$\\
    $\eta$ & Local learning rate\\
%    $\Theta^{(k)}$ & A mini-batch of samples for device $k$\\
    $\tau$ & Intra-cluster aggregation period\\
    $q \tau$ & Inter-cluster aggregation period\\
    $\mathcal{N}_i$ & Set of neighbors of edge server $i$\\
    $\mathbf{H}$ & Mixing matrix\\
    % $\pi$ & Number of gossip steps per round\\
    $\zeta$ & Second largest eigenvalue of $\mathbf{H}$\\
    % $G^2$ & Bounded gradients\\
    $\mu_n$ & Computing time of one local iteration for device $n$\\
    $\nu_n$ & Uploading time of one full model for device $n$\\
    $\alpha_n$ & Computing energy of one mini-batch SGD for device $n$\\
    $p_n$ & Transmission power of device $n$\\
    $\Tilde{\mathcal{T}}$ & Total time budget\\
    $\Tilde{\mathcal{E}}$ & Total energy budget\\
  \bottomrule
\end{tabular}
% \vspace{-0.5cm}
\end{table}

FL over mobile edge networks encounters significant challenges such as high training latency and energy consumption. To improve communication efficiency, research efforts have been made to reduce the size of communication data via model compression~\cite{wang2018atomo,sonee2021wireless,shlezinger2020uveqfed,reisizadeh2020fedpaq,han2020adaptivegrad,cui2022optimal,li2021talk,zhang2023communication}. An orthogonal research direction is to optimize the local update frequency~\cite{luo2021cost,wang2019adaptive} based on the observation that when the local update frequency increase, the number of communication rounds between devices and the central server may be reduced. Moreover, some recent studies~\cite{nori2021fast,xu2022adaptive} jointly optimize local updating frequency and model compression ratio to speed up training and reduce resource consumption. However, Nori et al.~\cite{nori2021fast} applied the identical local update frequency and model compression ratio to all heterogeneous devices, which limits the full utilization of each device's capacity. Xu et al.~\cite{xu2022adaptive} used an adaptive local update frequency and model compression ratio for heterogeneous devices, but their approach, like the aforementioned works~\cite{luo2021cost,wang2019adaptive,wang2018atomo,sonee2021wireless,shlezinger2020uveqfed,reisizadeh2020fedpaq,han2020adaptivegrad,zhang2023communication,cui2022optimal,li2021talk,nori2021fast,xu2022adaptive}, assume a central server that aggregates model updates from all edge devices and hence are not directly applicable to the CFEL system considered in this paper.

To achieve a scalable and communication-efficient FL system, recent works~\cite{castiglia2021multi,sun2023semi,zhong2021p,zhang2022scalable} considered CFEL over mobile edge networks, where multiple edge servers coordinate their associated subsets of devices independently and communicate with neighboring servers without relying on a central server. Specifically, Zhong et al.~\cite{zhong2021p} developed an algorithm named P-FedAvg under a two-tier communication network. P-FedAvg~\cite{zhong2021p} leverages multiple servers to significantly reduce the communication cost in FL. However, it did not take into account the system heterogeneity. 
Castiglia el al.~\cite{castiglia2021multi} and Sun et al.~\cite{sun2023semi} considered the device heterogeneity by allowing adaptive local update frequency for each device, but they did not optimize the local update frequency under resource constraints. Moreover, they assume full model transmission during training, which cannot address the straggler problem brought by the communication heterogeneity. Therefore, considering the system heterogeneity and dynamic system state, this paper aims to develop a novel scheme that adaptively determines the personalized local update frequency and compression ratio for each device at each round in CFEL to maximize model accuracy while minimizing training latency and energy consumption.

%%%%%%%%%%%%%%%%%%%%%%%%%%%%%%%%%%%%%%%%%%%%%%%%
\section{Preliminaries and Problem Formulation}\label{sec:prel_prob}
%%%%%%%%%%%%%%%%%%%%%%%%%%%%%%%%%%%%%%%%%%%%%%%%

In this section, we first introduce the CFEL network architecture and the existing learning algorithms. Then, we propose our algorithm to address the system heterogeneity and dynamic state in CFEL. Finally, we analyze the time and energy models and formulate our problem. 

%%%%%%%%%%%%%%%%%%%%%%%%%%%%%%%%%%%%%%%%%%%%%%%%%%%%%%%%
\subsection{CFEL: Cooperative Federated Edge Learning}\label{subsec_cfel}
\begin{figure}[t]
\centering
\includegraphics[width=0.47\textwidth]{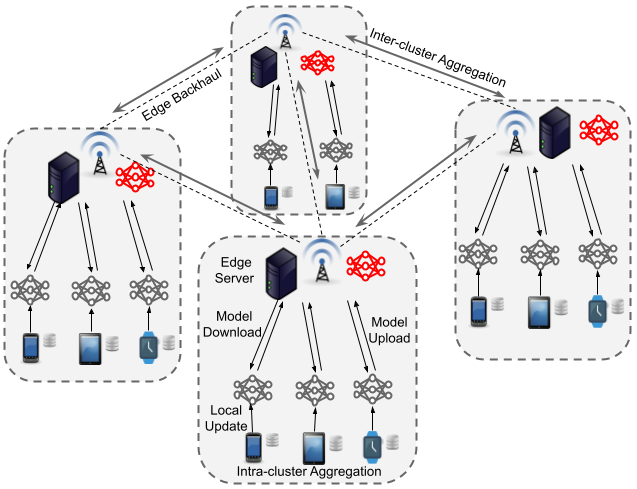}
\caption{CFEL: Cooperative Federated Edge Learning.}\label{fig:system}
% \vspace{-0.5cm}   
\end{figure}

%%%%%%%%%%%%%%%%%%%%%%%%%%%%%%%%%%%%%%%%%%%%%%%%%%%%%%%%
As shown in Fig.~\ref{fig:system}, we consider a CFEL system consisting of $N$ devices that are distributed among $m$ clusters. Each cluster $i\in[m]$ contains one edge server and a set of devices $\mathcal{S}_i$ with $N_i = |\mathcal{S}_i|$. A device $n\in\mathcal{S}_i$ is associated with edge server $i$ by some predefined criteria such as physical distance and wireless network protocols. The set of all edge devices in the system is denoted by $\mathcal{S}=\cup_{i=1}^{m}\mathcal{S}_i$ with $N=|\mathcal{S}|$. The edge backhaul network facilitates communication between edge servers. The topology of the edge backhaul network is represented as a connected and undirected graph $\mathcal{G}=\{\mathcal{V}, \mathcal{E}\}$, where $\mathcal{V}$ is the set of edge servers and $\mathcal{E}$ is the set of communication links between edge servers. Some main notations used in the paper are summarized in Table~\ref{tab:notations}.

The goal of FL is to find a global model $\mathbf{x} \in \mathbb{R}^d$ that minimizes the following optimization problem: 
\begin{equation}\label{prob:overall}
\min_{\mathbf{x}} F(\mathbf{x}) = \frac{1}{N}\sum_{n=1}^{N} F_n(\mathbf{x}),
\end{equation}
where $F_n(\mathbf{x}) = \mathbb{E}_{z \sim \mathcal{D}_n}[\ell_n(\mathbf{x}; z)]$ is the local objective function of device $n$, and $\mathcal{D}_n$ is the data distribution of device $n$. Here $\ell_n$ is the loss function defined by the learning task, and $z$ represents a data sample from the distribution $\mathcal{D}_n$. 

In our CFEL system, to achieve this goal, we decompose the global optimization problem into multiple subproblems, each corresponding to a cluster. The local objective function of the $i$-th cluster is defined as
\begin{equation}\label{prob:edge}
\min_{\mathbf{x}} f_i(\mathbf{x}) = \frac{1}{N_i}\sum_{n \in \mathcal{S}_i}  F_n(\mathbf{x}),
\end{equation}
which represents the average loss over all devices in cluster $i$. Then the global objective function~\eqref{prob:overall} can be reformulated as:
\begin{equation}\label{prob:global_edge}
\min_{\mathbf{x}} F(\mathbf{x}) = \sum_{i=1}^m \frac{N_i}{N} f_i(\mathbf{x}).
\end{equation} 
In CFEL, devices within the system collaboratively solve the above optimization problem under the coordination of the edge servers in their clusters without sharing the raw data.

The classic FedAvg algorithm cannot be directly applied to the CFEL system due to the presence of multiple clusters and lack of a central server for model aggregation. To solve \eqref{prob:overall} under the CFEL network architecture, new learning algorithms (e.g., MLL-SGD~\cite{castiglia2021multi}, SD-FEEL~\cite{sun2023semi}, P-FedAvg~\cite{zhong2021p} and CE-FedAvg~\cite{zhang2022scalable}) have been proposed. These algorithms mainly comprise three key stages as follows. (i) Local model update: each device performs multiple local iterations to update its local model via the mini-batch SGD; (ii) Intra-cluster aggregation: after updating the local model, each device uploads its updated local model to the corresponding edge server, and the edge server computes the averaged edger server model and send it back to the associated devices; and (iii) Inter-cluster aggregation: after multiple rounds of intra-cluster aggregations, each edge server communicates with its neighboring edge servers to do the inter-cluster aggregation, and sends the intra-cluster aggregated model to its associated devices. %
Although prior work has demonstrated the potential of such a new learning algorithm, how to address challenges such as system heterogeneity and
dynamic state in CFEL remains unknown. 

%----------------------------------------------------
%----------------------------------------------------

%%%%%%%%%%%%%%%%%%%%%%%%%
\subsection{Heterogeneity-Aware CFEL}\label{sub_sec:hete_cfel}
%%%%%%%%%%%%%%%%%%%%%%%%%

%-----------------------------------------------------
\begin{algorithm}[t]
\caption{Proposed HCEF Scheme.}\label{algorithm-1}
\begin{algorithmic}[1]
    \STATE Initialization: initial edge models $\mathbf{y}_{i}^{0, 0}$, $\forall i \in [m]$, edge backhaul graph $\mathcal{G}$, mixing matrix $\mathbf{H} \in [0, 1]^{m \times m}$, intra-cluster aggregation period $\tau$, and inter-cluster aggregation period $q\tau$. 
    \FOR{each global round $l = 0, \ldots, \phi - 1$}
        \FOR{each cluster $i \in [m]$ \textbf{in parallel}}
            \FOR{each edge round $r = 0, \ldots, q - 1$}
                \STATE Each edge server $i$ sends $\mathbf{y}_{i}^{l, r}$ to its associated devices $n\in\mathcal{S}_i$.\label{alg1:send}
                \STATE Each edge device $n$ uploads its local parameters $(\sigma_{n}^{l,r})^2$, $(G_{n}^{l,r})^2$, $\mu_{n}^{l,r}$, $\alpha_{n}^{l,r}$, and $\nu_{n}^{l,r}$ to the coordinator by Algorithm~\ref{algorithm-2}.\label{alg1:upload_parameters}
                    \STATE Coordinator finds the optimal compression ratios $\theta_n^{l,r}$ and probabilities $\rho_n^{l,r}$ for $n \in \mathcal{S}_i$ according to Algorithm~\ref{algorithm-3} and sends them to $\mathcal{S}_i$.\label{alg1:rho_theta}
                \FOR{each device $n \in \mathcal{S}_i$ \textbf{in parallel}}
                    
                    \STATE $\mathbf{x}_{n}^{l, r, 0} \gets \mathbf{y}_{i}^{l, r}$ \label{li:broadcast}
                    
                    \FOR{each local iteration $s = 0, \ldots, \tau - 1$}
                        \STATE Compute a mini-batch gradient $\mathbf{g}_n^{l,r,s}$ with probability $\rho_n^{l,r}$.\label{comp_sgd}
                        \STATE $\mathbf{x}_{n}^{l, r, s + 1} \gets \mathbf{x}_{n}^{l, r, s} - \eta \mathbf{g}_n^{l,r,s}$ \label{local_upd} 
                    \ENDFOR
                    \STATE $\Delta_{n}^{l,r}\gets Q(\mathbf{x}_{n}^{l, r, \tau} - \mathbf{x}_{n}^{l, r, 0})$\label{local_compress}
                \ENDFOR
                
             \STATE $\mathbf{y}_{i}^{l, r + 1} \gets \mathbf{y}_{i}^{l, r }+\frac{1}{N_i}\sum_{n \in \mathcal{S}_i}\Delta_{n}^{l,r}$ \label{li:ave_intra}%\hfill $\rhd$ Intra-cluster aggregation 
            \ENDFOR
            \STATE $\mathbf{y}_{i}^{l + 1, 0} \gets\sum_{j \in \{i\}\cup\mathcal{N}_i}\mathbf{H}_{j, i} \mathbf{y}_{j}^{l, q}$\label{li:ave_gossip}%\hfill $\rhd$ Inter-cluster aggregation
        \ENDFOR
    \ENDFOR
\end{algorithmic}
\end{algorithm}
%-----------------------------------------------------

To address the aforementioned challenges in CFEL, we present a new scheme called HCEF, which allows adaptive control of local update frequency and gradient compression. To monitor the dynamic state and heterogenous devices' computing and communication capabilities during the training, similar to prior decentralized FL works~\cite{wang2019matcha,wang2022accelerating,xu2021decentralized}, we assume that there exists a coordinator and ignore the cost (e.g., bandwidth consumption and time cost) for information collection since the size of this information (e.g., 100-300KB~\cite{lyu2018multi}) is much smaller than that of model parameters. Furthermore, the role of the coordinator can be dynamically assigned to different edge servers. This means that if one coordinator fails, other edge servers can quickly take over the coordination tasks, ensuring that the training process continues without interruption. We summarize the HCEF scheme for CFEL in Algorithm~\ref{algorithm-1}. The overall training process of HCEF is divided into $\phi$ global rounds wherein each cluster first performs $q$ edge rounds of intra-cluster collaboration independently and then communicates with other clusters for inter-cluster collaboration. 

Specifically, at the beginning of the $r$-th edge round in the $l$-th global round, denoted as the $(r,l)$-th round, each edge server $i$ broadcasts its edge model $\mathbf{y}_{i}^{l,r}$ to the associated devices $n\in\mathcal{S}_i$ (line~\ref{alg1:send}). After that, each device $n$ needs to estimate and upload several parameters to the coordinator, including the training parameters $(\sigma_{n}^{l,r})^2$, $(G_{n}^{l,r})^2$, computing parameters $\mu_{n}^{l,r}$, $\alpha_{n}^{l,r}$, and communication parameter $\nu_{n}^{l,r}$ (line~\ref{alg1:upload_parameters}). This process will be elaborated in Section~\ref{subsec_joint_iter}. Then, the coordinator uses Algorithm 3 to determine the compression ratio $\theta_{n}^{l,r}$ and local update probability $\rho_{n}^{l,r}$ for each device by the uploaded training parameters. This procedure will be elaborated in Section~\ref{sec:alg_design}.

Next, considering the heterogeneous computation capabilities of the devices, each device $n$ performs its updates independently and potentially asynchronously. The updating rule for each device can be described by the following:
\begin{gather}\label{eq:localSGD}
\mathbf{x}_{n}^{l,r,s+1} \gets \mathbf{x}_{n}^{l,r,s} - \eta \mathbf{g}_{n}^{l,r,s}, \forall s = 0, \ldots, \tau - 1,
\end{gather}
where $\eta$ is the local learning rate, and $\mathbf{g}_{n}^{l,r,s} $ is a random stochastic gradient determined by the adaptive local update probability (line~\ref{comp_sgd}-\ref{local_upd}).

After $\tau$ local iterations, each device compresses the model updates using the compression operator $Q(\cdot)$ according to the compression ratios $\theta_n^{l,r}=k_n^{l,r}/d$ where $k_n^{l,r}$ is the number of uploaded parameters for device $n$ at the $(r,l)-$th round (line~\ref{local_compress}) and sends the compressed updates $\Delta_{n}^{l,r}$ to the associated edge server $i$ for intra-cluster aggregation (line~\ref{li:ave_intra}). Specifically, once receiving all compressed model updates from the associated devices, each edge server $i$ updates its edge model $\mathbf{y}_{i}^{l, r}$ by aggregating the compressed model updates. It’s crucial to recognize that the compression ratio $\theta_n^{l,r}=k_n^{l,r}/d$ varies across devices, influenced by factors such as each device’s bandwidth resources, communication capabilities, and the quality of the model being updated. Higher compression ratios, while requiring more bandwidth resources and extending communication time, can enhance model quality and increase training accuracy. Conversely, lower ratios deteriorate model quality but conserve bandwidth resources and improve communication efficiency. Therefore, selecting the optimal compression ratio requires a careful balance, considering the specific bandwidth limitations, communication needs, and model quality requirements of each device.

After $q$ edge rounds, the edge servers engage in inter-cluster aggregation by communicating with their neighboring servers via the edge backhaul (line~\ref{li:ave_gossip}). Specifically, each edge server updates its model by taking the average with neighboring servers using gossip protocol as follows:
\begin{equation}\label{eq:ave_gossip}
\mathbf{y}_{i}^{l + 1, 0} \gets\sum_{j \in \{i\}\cup\mathcal{N}_i}\mathbf{H}_{j, i} \mathbf{y}_{j}^{l, q},
\end{equation}
where $\mathcal{N}_i = \{j: (j, i) \in \mathcal{E}\}$ represents the neighboring set of edge server $i$ in the graph $\mathcal{G}$, and $\mathbf{H} \in [0, 1]^{m \times m}$ denotes the mixing matrix. Each element $\mathbf{H}_{j, i}$ signifies the weight assigned by server $i$ to server $j$ and $\mathbf{H}_{j, i} > 0$ if and only if servers $i$ and $j$ are directly connected in the edge backhaul. Finally, the algorithm proceeds to the subsequent global round $l + 1$ until a total of $\phi$ global rounds have been completed.

In the following, we elaborate on the two primary innovations of our scheme, namely adaptive local update frequency and gradient compression.

\textbf{Adaptive Local Update Frequency.} Unlike traditional FL approaches where all devices update their local models at a fixed frequency, adaptive local update frequency allows devices to adapt their update frequency based on heterogeneous computing capabilities and varying network conditions in the FL system. 

In the heterogeneous FL system, the straggler may delay the training time and some devices may consume more energy than others using the same local update frequency. Therefore, the HCEF scheme dynamically allocates the local update probability to achieve different local update frequencies for edge devices to address the system heterogeneity. Specifically, at the $s$-th local iteration of $(r,l)$-th round, the random stochastic gradient $\mathbf{g}_{n}^{l,r,s}$ is determined by the local update probability $\rho_n^{l,r}$ as follows:
\begin{equation}
\mathbf{g}_{n}^{l,r,s}= 
\begin{cases}
\mathbf{g}_n(\mathbf{x}_n^{l,r,s}), & \text{with probability } \rho_n^{l,r}\\
    \bm{0}, & \text{with probability}   1-\rho_n^{l,r}
\end{cases}
\end{equation}
where $\mathbf{g}_n(\mathbf{x}_n^{l,r,s})$ is the stochastic gradient computed over a mini-batch sampled from the local data distribution.

In general, we will assign a higher local update probability to devices with larger computation capabilities in each round so that the straggler effect is mitigated. This will make our CFEL system more time and energy-efficient. 

\textbf{Adaptive Gradient Compression.} Gradient compression plays a crucial role in addressing the communication challenges and resource limitations in FL. Transmitting models or model updates between edge devices and the edge server can be a bottleneck due to limited bandwidth and high communication costs. By compressing the local updates before transmission, the communication overhead and latency can be significantly reduced, enabling more efficient and scalable FL systems. In this paper, we utilize $\topk_k$ sparsification as our compression strategy, which only transmits $k$ coordinates with the largest magnitude. The $\topk_k$ compression operator $Q$ with compression ratio $\theta_{n}^{l,r}$ satisfies the following contraction property~\cite{stich2018sparsified}:
\begin{equation}\label{eq_topk}
\textstyle\mathbb{E}\|Q(\mathbf{x}) - \mathbf{x}\|^2 \leq (1-\theta_{n}^{l,r})\|\mathbf{x}\|^2, \forall \mathbf{x}\in\mathbb{R}^d.
\end{equation}
Here, the 
It is worth noting that HCEF is also compatible with other sparsification methods, such as $\text{random}_k$~\cite{hu2023federated} and Johnson-Lindenstrauss (JL) random projection~\cite{sonee2021wireless}. Here, the compression error bound presented in Equation~\eqref{eq_topk} is applicable for both $\text{random}_k$ and $\text{top}_k$~\cite{stich2018sparsified}. In~\cite{m2021efficient}, the authors assume gradients follow a power law decay distribution and derive the compression error accordingly. However, we use the compression error bound from~\cite{stich2018sparsified} because it is more general, does not rely on specific gradient distributions, and avoids the complexities introduced by unknown hyper-parameters, ensuring better adaptability and robustness across different scenarios. Intuitively, devices with faster network connections can compress less (i.e., a larger value of $\theta_n^{l, r}$) in each round, allowing them to transmit more meaningful model parameters, while devices with slower network connections will transmit less information. 

Considering the heterogeneous computing and communication capabilities of devices, HCEF dynamically adjusts the local update frequency and compression ratio for each device across rounds. This joint optimization is designed to improve both time and energy efficiency of CFEL, especially in the heterogeneous setting.

%%%%%%%%%%%%%%%%%%%%%%%%%%%%%%%%%%%%%%%%%%
\subsection{Problem Formulation}\label{sub_prob_form}
%%%%%%%%%%%%%%%%%%%%%%%%%%%%%%%%%%%%%%%%%%
Our goal is to minimize the training loss while satisfying the time and energy constraints via the joint control on computation probabilities $\{\rho_{n}^{l,r}\}$ and compression ratios $\{{\theta}_{n}^{l,r}\}$ in CFEL. To formulate our optimization problem, we first introduce the time consumption model and energy consumption model. These models will help us understand how the time and energy consumption of our system vary depending on the control variables. 

\textbf{Time Consumption Model.} We consider both computation and communication time. Since the download bandwidth is typically much higher than the upload bandwidth~\cite{kairouz2021advances}, we only focus on the uploading time in the system modeling without loss of generality as in \cite{xu2022adaptive,jiang2023heterogeneity,wang2022accelerating,zhang2022scalable}. We define $\mu_{n}^{l,r}$ and $\nu_{n}^{l,r}$ as the computing time for one local iteration and upload time of one full model for device $n$ at $(r,l)$-th round, respectively, and $\mathcal{T}_{i,i^{\prime}}^{l}$ as the transmission time from cluster $i$ to its neighbor $i^\prime$. Taking into account the heterogeneous local update probability $\rho_{n}^{l,r}$ and compression ratio $\theta_{n}^{l,r}$, the total expected training time for $\phi$ global rounds can be determined as:
\begin{equation}\label{eq:time}
\mathcal{T}= \sum_{l=0}^{\phi-1}\max_{i\in[m]}\{\sum_{r=0}^{q-1}\max_{n\in\mathcal{S}_i}\{\rho_n^{l,r}\tau\mu_n^{l,r} + \theta_{n}^{l,r}\nu_n^{l,r} \} + \max_{i^\prime\in\mathcal{N}_{i}}\mathcal{T}_{i,i^{\prime}}^{l}\},
\end{equation}
where $\rho_n^{l,r}\tau\mu_n^{l,r}$ and $\theta_{n}^{l,r}\nu_n^{l,r}$ are the computing time and communication time of device $n$ at $(r,l)$-th round. The completion time of $l$-th global round is determined by the ``slowest" cluster $i\in[m]$, whose completion time is determined by the ``slowest" device $n\in\mathcal{S}_i$ and the communication time to its neighbors.

\textbf{Energy Consumption Model.} The energy consumption includes both computing energy and communication energy. We define the one step of mini-batch SGD energy and transmission power of device $n$ at the $(r,l)$-th round as $\alpha_{n}^{l,r}$ and $p_{n}$. Following~\cite{yang2020energy}, considering the heterogeneous local updating probability $\rho_{n}^{l,r}$ and compression ratio $\theta_{n}^{l,r}$ for each device $n$, the expected total energy for $\phi$ global rounds can be determined as:
\begin{equation}\label{eq:energy}
    \mathcal{E} =\sum_{l=0}^{\phi-1}\sum_{r=0}^{q-1}\sum_{n=1}^{N}(\rho_n^{l,r}\tau\alpha_{n}^{l,r} +  p_{n}\theta_{n}^{l,r} \nu_n^{l,r}),
\end{equation}
where $\rho_n^{l,r}\tau\alpha_{n}^{l,r}$ and $p_{n}\theta_{n}^{l,r} \nu_n^{l,r}$ are the computing energy and communication energy of device $n$ at $(r,l)$-th round, respectively. The total energy consumption is the sum of energy consumed by all devices over all edge rounds.

Note that we do not have a global model in CFEL like the traditional FL. Instead, we focus on the averaged model across devices at the $(l, r)$-th round defined as
\begin{equation}\label{eq:aver_model}
\mathbf{u}^{l,r} = \frac{1}{N}\sum_{n=1}^{N}\mathbf{x}_n^{l,r}. 
\end{equation}
Here, the averaged model is widely used in the convergence analysis of decentralized algorithms in the literature~\cite{nedic2018network,sun2023semi,zhang2022scalable}. In practice, the edge models will reach consensus after convergence, and the averaged model will become the single global model. With the above time and energy models, our problem can be formulated as the following constrained optimization problem:
\begin{subequations}
\begin{align}
    \textbf{P1:} \quad &\min_{\{\rho_n^{l,r}\},\{\theta_{n}^{l,r}\}} & &\mathbb{E} [F(\mathbf{u}^{\phi,q})]\\
    &\text{s.t.} & & \mathcal{T}\leq\Tilde{\mathcal{T}}\label{cons:time_ori}\\
    & & & \mathcal{E}\leq\Tilde{\mathcal{E}}\label{cons:energy_ori}\\
    & & & 0 < \rho_{n}^{l,r} \leq 1, \quad \forall n, r, l\label{cons:prob}\\
    & & & 0 < \theta_{n}^{l,r} \leq 1, \quad \forall n, r, l,\label{cons:frequency}
\end{align}
\end{subequations}
where $\Tilde{\mathcal{T}}$ and $\Tilde{\mathcal{E}}$ are the total time and energy budgets, respectively.

Solving \textbf{P1} faces several challenges. First, there is no explicit mathematical expression to capture the relationship between $\{\rho_n^{l,r}\},\{\theta_n^{l,r}\}$ and $F(\mathbf{u}^{\phi,q})$. To address it, we analyze the convergence properties of the learning algorithm in Section~\ref{sec_conv} and then use the convergence error bound to substitute the loss function as the surrogate objective in Section~\ref{sec:alg_design}. Second, the time constraint~\eqref{cons:time_ori} and energy constraint~\eqref{cons:energy_ori} depend on the statuses of the overall training process, which are usually random. Thus, we develop an online algorithm to solve \textbf{P1} without requiring prior knowledge of future system information in Section~\ref{sec:alg_design}.

%%%%%%%%%%%%%%%%%%%%%%%%%%%%%%%%%%%%%%
\section{Convergence Analysis}\label{sec_conv}
%%%%%%%%%%%%%%%%%%%%%%%%%%%%%%%%%%%%%%
In this section, we give the convergence properties of Algorithm \ref{algorithm-1} under general non-convex settings. Before stating our convergence results, we make the following assumptions:

\begin{assumption}[Smoothness]\label{ass:smoothness}
Each local objective function $F_n:\mathbb{R}^d\rightarrow\mathbb{R}$ is $L$-smooth for all $n\in [N]$, i.e.,
\[
    \|\nabla F_n(\mathbf{x}) - \nabla F_n(\mathbf{x}^\prime)\| \leq L \|\mathbf{x} - \mathbf{x}^{\prime}\|,    \; \forall \mathbf{x},\mathbf{x}^{\prime}\in \mathbb{R}^d. 
\]
\end{assumption}

\begin{assumption}[Unbiased Gradient and Bounded Variance]\label{ass:gradient}
The local mini-batch stochastic gradient is an unbiased estimator of the local gradient: $\mathbb{E}[\mathbf{g}_n(\mathbf{x})] = \nabla F_n(\mathbf{x})$ and has bounded variance: $\mathbb{E}[\|\mathbf{g}_n(\mathbf{x}) - \nabla F_n(\mathbf{x})\|^2] \leq \sigma^2, \forall \mathbf{x} \in \mathbb{R}^d, n \in [N]$.
\end{assumption}

\begin{assumption}[Lower Bounded]\label{ass:lowerbounded}
 There exists a constant $F_{\inf}$ such that
\[
F(\mathbf{x}) \geq F_{\inf}, \forall \mathbf{x}\in\mathbb{R}^d.
\]
\end{assumption}
\begin{assumption}[Bounded gradients]\label{ass:bounded_grad}
There exists a constant $G\geq0$ such that
\[
\|\nabla F_n(\mathbf{x})\|^2 \leq G^2, \forall \mathbf{x} \in \mathbb{R}^d, n \in [N].
\]
\end{assumption}

\begin{assumption}[Mixing Matrix]\label{ass:mixing}
The graph $\mathcal{G}: = (\mathcal{V}, \mathcal{E})$ is strongly connected and the mixing matrix $\mathbf{H} \in [0, 1]^{m \times m}$ defined on it satisfies the following:
\begin{enumerate}
    \item $(i, j) \in \mathcal{E}$, then $\mathbf{H}_{i, j} > 0$; otherwise, $\mathbf{H}_{i, j} = 0$. 
    
    \item $\mathbf{H}$ is symmetric doubly stochastic, i.e., $\mathbf{H}\bm{1}= \bm{1}, \bm{1}^\intercal\mathbf{H}=\bm{1}^\intercal$.
    
    \item The magnitudes of all eigenvalues except the largest one are strictly less than 1, i.e., $\zeta =  \max \{{|\lambda_2(\mathbf{H})|,|\lambda_m(\mathbf{H})|}\}<\lambda_1(\mathbf{H})=1$.
\end{enumerate}
\end{assumption}
%-------------------------------

Assumptions~\ref{ass:smoothness},~\ref{ass:gradient},~\ref{ass:lowerbounded}, and~\ref{ass:bounded_grad} are standard in the analysis of SGD \cite{bottou2018optimization}, \cite{guo2022hybrid}. Assumption~\ref{ass:mixing} follows the decentralized optimization literature \cite{koloskova2020unified,zhang2022scalable} and ensures that the gossip step converges to the average of all the vectors shared between the nodes in the graph $\mathcal{G}$. Here, smaller $\zeta$ indicates better connectivity between edge servers. For example, for complete graphs and bipartite graphs, $\zeta=0$ and $\zeta=1$, respectively.

Next, we propose our main theoretical result of the HCEF algorithm in the following theorem. 
For the convenience of mathematical derivation, we define $t = l q\tau+r\tau+s$, where $ l \in [0, \phi-1]$, $r \in [0, q-1]$ and $s \in [0, \tau-1]$, as the global iteration index, and $T= \phi q\tau$ as the total number of global training iterations in Algorithm~\ref{algorithm-1}.

%%%%%%%%%%%%%%%%%%%%%%%%%%%%%%%%%%%
\begin{lemma}[Convergence Decomposition]~\label{lemma_bound_ave_grad} Under Assumptions~\ref{ass:smoothness}, \ref{ass:gradient}, and \ref{ass:lowerbounded}, if the learning rate $\eta \leq  (\rho_n^{l,r})^2-2(\rho_n^{l,r})^2(1-\theta_n^{l,r})\/(2L(2-\theta_n^{l,r})\rho_{n}^{l,r}), \forall n, r, l,$ the iterates of Algorithm~\ref{algorithm-1} satisfy 
\begin{align}
 &\frac{1}{T}\sum_{t=0}^{T-1}\mathbb{E}\|\nabla F(\mathbf{u}^{t})\|^2 
 \leq \frac{12}{NT}\sum_{t=0}^{T-1}\sum_{n=1}^N(1-\rho_n^t)^2\mathbb{E}\|\nabla F_n(\mathbf{x}_n^{t})\|^2  \notag\\
 &+\frac{4\eta L}{NT}\sum_{t=0}^{T-1}\sum_{n=1}^{N}(2-\theta_{n}^t)\rho_n^t\sigma^2+\frac{4( F(\mathbf{u}^0)- F_{\inf})}{\eta T}\notag\\
 & +\frac{4L^2}{NT}\sum_{t=0}^{T-1}\sum_{n=1}^{N}\mathbb{E}\|\mathbf{u}^{t}-\mathbf{x}_n^{t}\|^2.\label{lemma1_upp}
\end{align}
\end{lemma}
%%%%%%%%%%%%%%%%%%%%%%%%%%%%%%%%%%%%%
\begin{IEEEproof}
The proof is provided in Appendix~\ref{append_lemma} in the supplementary text.
\end{IEEEproof}

Lemma~\ref{lemma_bound_ave_grad} aims to provide the composition of the total convergence error bound. The convergence bound~\eqref{lemma1_upp} contains two parts. The first three terms represent optimization errors resulting from SGD under the compression and update probability. When no compression and update probability is applied (i.e., $\theta_n^{t}=1$ and $\rho_n^t=1$), the error corresponds to the \textit{Fully synchronous SGD} error in~\cite{wang2021cooperative}. The last term $\|\mathbf{u}^{t}-\mathbf{x}_n^{t}\|^2$ represents the discrepancy error between the device models $\mathbf{x}_n^{t}$ and the global average model $\mathbf{u}^{t}$. Next, to establish the full convergence of Algorithm~\ref{algorithm-1}, we provide the upper bounds for the discrepancy error. 

%--------------------------------------------
\begin{lemma}[Bounded Discrepancy Error]~\label{lemma_bound_disc_error} Under Assumptions~\ref{ass:gradient} and~\ref{ass:mixing}, we have
\begin{align*}
% \begin{equation}
    \sum_{t=0}^{T-1}\sum_{n=1}^N\mathbb{E}\|\mathbf{u}^{t}-\mathbf{x}_{n}^{t}\|^2 \leq & 4\eta^2q^2\tau^2 \Omega_{1} \sum_{t=0}^{T-1}\sum_{n=1}^{N}\big[(2-\theta_{n}^t)\rho_n^t\sigma^2 \\
    &+ (2-\theta_{n}^t)\rho_n^t\|\nabla F_n(\mathbf{x}_n^{t})\|^2\big].
% \end{equation}
\end{align*} 
\end{lemma}
%--------------------------------------------
\begin{IEEEproof}
    The proof is presented in Appendix~\ref{append_lemma_bound_disc}.
\end{IEEEproof}

Lemma~\ref{lemma_bound_disc_error} gives the upper bound of the discrepancy error term. Combining Lemmas~\ref{lemma_bound_ave_grad},~\ref{lemma_bound_disc_error} and choosing a proper learning rate, we arrive at the following final convergence bound.
%----------------------------------------------
\begin{theorem}[Convergence of HCEF]\label{th:convergence}
Let Assumptions~\ref{ass:smoothness}--\ref{ass:mixing} hold, and let $L$, $\sigma$, $G$ be as defined therein. If the learning rate satisfies 
\begin{equation*}
    \eta \leq \min\{\frac{1}{4Lq^2\tau^2\Omega_1}, \frac{(\rho_n^{l,r})^2-2(\rho_n^{l,r})^2(1-\theta_n^{l,r})}{2L(2-\theta_n^{l,r})\rho_{n}^{l,r}}, \forall l,r,n\},
\end{equation*}
where 
\begin{equation*}
    \Omega_{1}=\frac{1}{1-\zeta^{2}}+\frac{2}{1-\zeta}+\frac{\zeta}{(1-\zeta)^{2}},
\end{equation*} 
then for any $\phi q\tau > 0$, the iterates of Algorithm~\ref{algorithm-1} for HCEF satisfy
\begin{align}
\frac{1}{\phi q}&\sum_{l=0}^{\phi-1}\sum_{r=0}^{q-1}\mathbb{E}\|\nabla F(\mathbf{u}^{l,r})\|^2 \leq \frac{4( F(\mathbf{u}^{0,0})- F_{\inf})}{\eta \phi q\tau} \notag\\
& + \frac{8 \eta L(\sigma^2+G^2)}{N\phi q}\sum_{l=0}^{\phi-1}\sum_{r=0}^{q-1}\sum_{n=1}^{N}(2-\theta_n^{l,r})\rho_n^{l,r}\notag\\
 & +\frac{12G^2}{N\phi q}\sum_{l=0}^{\phi-1}\sum_{r=0}^{q-1}\sum_{n=1}^{N}(1-\rho_{n}^{l,r})^2.\label{eq_theo_g_sigma}
\end{align}
\end{theorem}
\begin{IEEEproof}
    The proof is presented in Appendix~\ref{append:theorem}.
\end{IEEEproof}
%-----------------------------------------------

The first term is identical to centralized SGD~\cite{bottou2018optimization}. As $\phi q\tau \rightarrow \infty$, this term goes to zero. The second term is the stochastic error related to the compression ratios and local update probabilities. If the stochastic gradients are highly diverged, the local update frequency is high, or the compression ratio is low, then the stochastic error will be more significant. In line with~\cite{castiglia2021multi}, we observe that when the updating probability is low, the stochastic error becomes smaller. The third term reveals the additive error depends on the local update probability. If there is no adaptive local update frequency, i.e., $\rho_{k}^{l,r}=1$, the error is equal to zero. Furthermore, Theorem 1 shows that the iteration complexity of HCEF is $\mathcal{O}(1/(\eta\phi q\tau)+\eta \sigma^2 + (\eta+1)G^2)$.

\begin{remark}[\textbf{Effect of $\theta_n^{l,r}$ and $\rho_n^{l,r}$}]\label{remark_effe_thet_rho} We analyze the impact of the compression and local update probability on the convergence of HCEF. First, higher values of \(\theta_n^{l,r}\), implying transmitting more parameters, effectively reduce the contribution of stochastic error by decreasing the term \(2 - \theta_n^{l,r}\). If $\theta_n^{l,r}$ is close to 1, it minimizes the contribution of the corresponding stochastic error term. Second, higher $\rho_n^{l,r}$ values, implying more frequent updates, increase the contribution of the stochastic error. However, it reduces the additive error term by decreasing the term \(1 - \rho_n^{l,r}\). Therefore, we need to carefully choose the $\rho_n^{l,r}$ to balance the convergence speed.
\end{remark}

\begin{remark}[\textbf{Comparison to FedAvg}]\label{remark_comp} When there is neither compression nor adaptive local updates, meaning that $\theta_n^{l,r}=\rho_n^{l,r}=1$ for all $n, l, r$, and all devices communicate with a single edge server with $q=1$, our HCEF algorithm reduces to the FedAvg algorithm. In this case, if the learning rate satisfies $\eta={1}/{(L\sqrt{T})}$ when $T>\tau^4$, the iteration complexity of HCEF satisfies $\mathcal{O}({1}/{\sqrt{T}})+\mathcal{O}({(\sigma^2+G^2)}/{\sqrt{T}}).$ This coincides with the complexity of FedAvg given in~\cite{yu2019parallel}.
\end{remark}

%%%%%%%%%%%%%%%%%%%%%%%%%%%%%%%%%
\section{Algorithm Design}\label{sec:alg_design}
%%%%%%%%%%%%%%%%%%%%%%%%%%%%%%%%%

In this section, we first reformulate Problem~\textbf{P1} using the convergence error bound in Theorem~\ref{th:convergence} and divide it into a series of one-slot problems. Then, we propose our solution algorithm to dynamically adjust the local update frequency and communication compression ratio under system heterogeneity and dynamic state.
%------------------------------------
\subsection{Problem Reformulation}\label{subsec:prob_refo}
%------------------------------------

For problem \textbf{P1}, the objective function can be substituted by the approximate upper bound in~\eqref{eq_theo_g_sigma} when $\eta\leq 3/(2L)$. Therefore, we obtain the following objective function after ignoring the constant terms:
\begin{equation}
    \sum_{l=0}^{\phi-1}\sum_{r=0}^{q-1}\sum_{n=1}^{N}[(2-\theta_n^{l,r})\rho_n^{l,r}(\sigma^2+G^2)+3(1-\rho_n^{l,r})^2G^2],
\end{equation}
which reveals the effects of $\rho_n^{l,r}$ and $\theta_{n}^{l,r}$ on the training process. Meanwhile, the coordinator in the system needs the entire training information (e.g., computing and communication parameters $\mu_{n}^{l,r}$, $\nu_{n}^{l,r}$, and $\alpha_{n}^{l,r}$) to solve \textbf{P1}. However, such parameters are usually time-varying and random caused by the dynamic system states. Thus, the coordinator cannot obtain such information in advance. To address the challenge of uncertainty, we leverage a greedy approach and tackle the multi-slot optimization problem by breaking it down into a series of one-slot problems. This method allows for real-time decision-making that accommodates the rapidly changing state of the system~\cite{jiang2023heterogeneity} and provides a robust framework against uncertainties in parameter estimations and system dynamics. Therefore, we solve the following optimization problem at each $(r,l)$-th round:

\begin{subequations}
\textbf{P2:}
\begin{align}
 % & \textbf{P2:}  & & \notag \\
&\min_{\bm{\rho}, \bm{\theta}}& & \sum_{n=1}^{N}[(2-\theta_n^{l,r})\rho_n^{l,r}(\sigma^2+G^2)+3(1-\rho_n^{l,r})^2G^2]\\
&\text{s.t.} & &(\phi-l)\max_{i\in[m]}\Big\{(q-r)\max_{n\in\mathcal{S}_i}\{\rho_n^{l,r}\tau\mu_n^{l,r} + \theta_{n}^{l,r}\nu_n^{l,r}\}\notag\\
& & & \quad+\sum_{e=0}^{r-1}\max_{n\in\mathcal{S}_i}\mathcal{T}_i^{l,e} + \max_{i^\prime\in\mathcal{N}_{i}}\mathcal{T}_{i,i^{\prime}}^{l}  \Big\}+\sum_{c=0}^{l-1}\mathcal{T}^{c}  \leq\Tilde{\mathcal{T}}\label{cons:time}\\
& & &  (\phi-l)\Big[(q-r)\sum_{n=1}^{N}(\rho_n^{l,r}\tau\alpha_{n}^{l,r} + p_{n}\theta_{n}^{l,r} \nu_n^{l,r}) \notag\\
& & & \quad+ \sum_{e=0}^{r-1}\mathcal{E}^{l,e} \Big] +\sum_{c=0}^{l-1}\mathcal{E}^{c}\leq\Tilde{\mathcal{E}}\label{cons:energy}\\
& & &  0 < \rho_{n}^{l,r} \leq 1, \quad \forall n\label{cons:probability}\\
& & &  0 < \theta_{n}^{l,r} \leq 1, \quad \forall n\label{cons:compression}
\end{align}
\end{subequations}
where $\bm{\rho} := \{\rho_{n}^{l,r}, \forall n\}$, $\bm{\theta} := \{\theta_{n}^{l,r}, \forall n\}$, $\sum_{c=0}^{l-1}\mathcal{T}^{c}$ represents the time consumed from the global round $0$ to $l-1$ and $\sum_{e=0}^{r-1}\max_{n\in\mathcal{S}_i}\mathcal{T}_i^{l,e} + \max_{i^\prime\in\mathcal{N}_{i}}\mathcal{T}_{i,i^{\prime}}^{l}$ denotes the time consumption from the $(0,l)$-th round to the $(r-1,l)$-th round and the communication time for the edge server $i$ at the $l$-th global round in the time constraint~\eqref{cons:time}, $\sum_{c=0}^{l-1}\mathcal{E}^{c}$ is the energy consumed from the global round 0 to $l-1$, and $ \sum_{e=0}^{r-1}\mathcal{E}^{l,e}$ represents the energy consumed from the $(0,l)$-th round to the $(r-1, l)$-th round for all devices.

Note that for the overall time constraint~\eqref{cons:time}, during the $l$-th global round, we utilize the time consumed in the $(r,l)$-th round to estimate the time consumption of the remaining $q - r$ edge rounds for the edge server $i$. Then, the remaining $\phi - l$ global round time can be further estimated by the current time consumption of the global round $l$. Similarly, we estimate the energy consumption of the remaining $q - r$ edge rounds by the energy consumed in the $(l, r)$-th round, and then the energy consumption of the remaining $\phi - l$ global rounds can be estimated by the current energy consumption of the global round $l$.

There are two primary challenges in solving Problem \textbf{P2}: (i) the coordinator is unaware of the learning parameters such as $\sigma$ and $G$, as well as local computing and transmission parameters such as $\mu_{n}^{l,r}$, $\nu_{n}^{l,r}$, and $\alpha_{n}^{l,r}$; and (ii) Problem \textbf{P2} is a non-convex optimization problem, which can be easily proved by checking the Hessian matrix of the objective function, and hence challenging to solve. Therefore, we develop an alternating minimization algorithm to get the optimal solution approximately for Problem~\textbf{P2} in the following.

%------------------------
\subsection{Algorithm Description}\label{subsec_joint_iter}
%------------------------

To tackle the aforementioned challenges, we develop a solution strategy where the devices (Algorithm~\ref{algorithm-2}) and the coordinator (Algorithm~\ref{algorithm-3}) collaboratively solve Problem \textbf{P2}. Specifically, at the beginning of each $(l, r)$-th round, each edge device $n\in\mathcal{S}_i$ first estimates the local unknown learning variables such as $\sigma_{n}^{l,r}$ and $G_{n}^{l,r}$ by the received edge model $\mathbf{y}_i^{l,r}$. Then, it estimates the computing parameters $\mu_{n}^{l,r}$, $\alpha_{n}^{l,r}$ and communication parameter $\nu_{n}^{l,r}$ according to the batch size, available CPU frequency and channel conditions between the device and server using models in~\cite{yang2020energy,tran2019federated,chen2020joint} without performing actual training. These parameters are then sent to the coordinator as shown in Algorithm~\ref{algorithm-2}. After receiving the parameters, the optimal solutions $\bm{\rho}^*$ and $\bm{\theta}^*$ of Problem \textbf{P2} can be efficiently computed using an iterative algorithm. The overall algorithm is given in Algorithm~\ref{algorithm-3}. Specifically, the coordinator first takes averages of parameters $(\sigma_n^{l,r})^2$, $(G_n^{l,r})^2$ to estimate $\sigma^2$ and $G^2$ (line 1-3). Next, the coordinator solves Problem \textbf{P2} iteratively by dividing it into two sub-problems.

On one hand, for a given $\bm{\rho}$, Problem \textbf{P2} becomes
\begin{align*}
\textbf{P2.1:} \quad &\min_{ \bm{\theta}} & &\sum_{n=1}^{N}(2-\theta_n^{l,r})\rho_n^{l,r}\\
&\text{s.t.} & &(\phi-l)\max_{i\in[m]}\Big\{(q-r)\max_{n\in\mathcal{S}_i}\{\rho_n^{l,r}\tau\mu_n^{l,r} + \theta_{n}^{l,r}\nu_n^{l,r}\}\notag\\
& & & \quad+\sum_{e=0}^{r-1}\max_{n\in\mathcal{S}_i}\mathcal{T}_i^{l,e} + \max_{i^\prime\in\mathcal{N}_{i}}\mathcal{T}_{i,i^{\prime}}^{l}  \Big\}+\sum_{c=0}^{l-1}\mathcal{T}^{c}  \leq\Tilde{\mathcal{T}}\\
& & &  (\phi-l)\Big[(q-r)\sum_{n=1}^{N}(\rho_n^{l,r}\tau\alpha_{n}^{l,r} + p_{n}\theta_{n}^{l,r} \nu_n^{l,r}) \notag\\
& & & \quad + \sum_{e=0}^{r-1}\mathcal{E}^{l,e}\Big]+\sum_{c=0}^{l-1}\mathcal{E}^{c} \leq\Tilde{\mathcal{E}}\label{cons:energy}\\
& & &  0 < \theta_{n}^{l,r} \leq 1, \quad \forall n.
\end{align*}
Problem \textbf{P2.1} is a linear program and can be easily solved. On the other hand, for a fixed $\bm{\theta}$, let $C_{n}^{l,r}=(2-\theta_n^{l,r})\sigma^2-(4+\theta_n^{l,r})G^2$. Problem \textbf{P2} becomes
\begin{align*}
\textbf{P2.2:} \quad & \min_{\bm{\rho}} &  &\sum_{n=1}^{N}[3(\rho_{n}^{l,r})^2G^2+\rho_{n}^{l,r}C_{n}^{l,r}]\\
&\text{s.t.} & &(\phi-l)\max_{i\in[m]}\Big\{(q-r)\max_{n\in\mathcal{S}_i}\{\rho_n^{l,r}\tau\mu_n^{l,r} + \theta_{n}^{l,r}\nu_n^{l,r}\}\notag\\
& & & \quad+\sum_{e=0}^{r-1}\max_{n\in\mathcal{S}_i}\mathcal{T}_i^{l,e} + \max_{i^\prime\in\mathcal{N}_{i}}\mathcal{T}_{i,i^{\prime}}^{l}  \Big\}+\sum_{c=0}^{l-1}\mathcal{T}^{c}  \leq\Tilde{\mathcal{T}}\\
& & &  (\phi-l)\Big[(q-r)\sum_{n=1}^{N}(\rho_n^{l,r}\tau\alpha_{n}^{l,r} + p_{n}\theta_{n}^{l,r} \nu_n^{l,r}) \notag\\
& & & \quad+ \sum_{e=0}^{r-1}\mathcal{E}^{l,e}\Big] +\sum_{c=0}^{l-1}\mathcal{E}^{c} \leq\Tilde{\mathcal{E}}\\
& & &  0 < \rho_{n}^{l,r} \leq 1, \quad \forall n.
\end{align*}
Problems \textbf{P2.2} is a quadratic programming problem and can be efficiently solved by a quadratic programming solver, e.g., CVX~\cite{diamond2016cvxpy}. In each iteration $e$, the coordinator finds the optimal value of $\bm{\theta}_{e+1}$ while keeping $\bm{\rho}_{e}$ fixed (line~\ref{alg3:theta}). Then $\bm{\rho}_{e+1}$ is updated with the obtained $\bm{\theta}_{e+1}$ in the preceding step (line~\ref{alg3:rho}). This iterative procedure continues until the convergence criteria $\epsilon$ are met or the maximum number of iterations $I_{\max}$ is reached.
% This iterative procedure continues until met convergence criteria $\epsilon$. 
Since Problems \textbf{P2.1} and \textbf{P2.1} are
both convex, both of them can be solved by using some known
optimization methods, e.g., interior point method.

\begin{remark}[Computational Complexity Analysis:]\label{remark_comp_se_feel} In our proposed Algorithm~\ref{algorithm-3}, the main problem \textbf{P2} is addressed by solving Problem \textbf{P2.1} alternatively
followed by Problem \textbf{P2.2}. According to~\cite{wang2014outage}, solving Problems \textbf{P2.1} and \textbf{P2.2} by interior point method are with the complexity order of $\mathcal{O}(N^{3.5})$ in each round. Denote the iteration number in Algorithm~\ref{algorithm-3} is $I_{iter}$. Then, the overall complexity of solving Problem \textbf{P2} is bounded by $\mathcal{O}(I_{iter}N^{3.5})$, which is bounded by $\mathcal{O}(I_{max}N^{3.5})$.
\end{remark}

%-----------------------------------------------------
\begin{algorithm}[t]
\caption{Procedure at device $n$}\label{algorithm-2}
\begin{algorithmic}[1]
    
    \STATE Estimate $(\sigma_{n}^{l,r})^2\leftarrow\mathbb{E}[\|\mathbf{g}_n(\mathbf{y}_{i}^{l,r})-\nabla F_n(\mathbf{y}_{i}^{l,r})\|^2]$\label{alg2:estimate_sigma}
    \STATE Estimate $(G_{n}^{l,r})^2\leftarrow\mathbb{E}[\|\nabla F_n(\mathbf{y}_{i}^{l,r})\|^2]$\label{alg2:estimate_G}
    \STATE Estimates $\mu_{n}^{l,r}$, $\alpha_{n}^{l,r}$, $\nu_{n}^{l,r}$\label{alg2:estimate_comp_comm}
    \STATE Uploads $(\sigma_{n}^{l,r})^2$, $(G_{n}^{l,r})^2$, computing parameters $\mu_{n}^{l,r}$, $\alpha_{n}^{l,r}$, communication parameter $\nu_{n}^{l,r}$ to the coordinator.\label{alg2:upload}
\end{algorithmic}
\end{algorithm}
%-----------------------------------------------------
%-----------------------------------------------------
\begin{algorithm}[t]
\textbf{Input:} Time and energy budgets $\tilde{\mathcal{T}}$ and $\tilde{\mathcal{E}}$, stopping criteria $\epsilon$ and $I_{max}$, $I=1$;\\
\textbf{Output:} $\bm{\theta}^* = \{\theta_n^{l,r},\forall n\}$, $\bm{\rho}^*=\{\rho_n^{l,r},\forall n\}$;\\
\caption{Procedure at the coordinator}\label{algorithm-3}
\begin{algorithmic}[1]
    \STATE Receives $\{(\sigma_{n}^{l,r})^2, \forall n\}$, $\{(G_{n}^{l,r})^2, \forall n\}$, $\{\mu_{n}^{l,r}, \forall n\}$, $\{\nu_{n}^{l,r}, \forall n\}$, $\{\alpha_{n}^{l,r}, \forall n\}$ from each device.\label{alg3_receive}
    \STATE $\sigma^2\leftarrow\frac{1}{N}\sum_{n=1}^{N}(\sigma_n^{l,r})^2$\label{alg3_estimate_sigma}
    \STATE $G^2\leftarrow\frac{1}{N}\sum_{n=1}^{N}(G_n^{l,r})^2$\label{alg3_estimate_G}
    \STATE $\bm{z}_0\leftarrow(\bm{\rho}_0 ,\bm{\theta}_0)$ and $e\leftarrow 0$ 
    \WHILE{ $\|\bm{z}_{e}-\bm{z}_{e-1}\|>\epsilon$ and $I<I_{max}$ }\label{alg3:criteria} 
    \STATE $\bm{\theta}_{e+1} \leftarrow$ solve Problem \textbf{P2.1} %\blue{(The index in P2.1 does not match)}\orange{explain the index e in main paper.}
    \label{alg3:theta}
    \STATE $\bm{\rho}_{e+1} \leftarrow$ solve Problem \textbf{P2.2} 
    %\blue{(The index in P2.2 does not match)}
    \label{alg3:rho}
    \STATE $I\leftarrow I+1$
    \ENDWHILE
    \STATE $\bm{\theta}^*,\bm{\rho}^*\leftarrow\bm{\theta}_{e+1},\bm{\rho}_{e+1}$ \label{alg3_update_theta_rho}
\end{algorithmic}
\end{algorithm}
%-----------------------------------------------------

%%%%%%%%%%%%%%%%%%%%%%%%%%%%%%%%%
\section{Performance Evaluation}\label{sec:exp}
%%%%%%%%%%%%%%%%%%%%%%%%%%%%%%%%%
%-------------------------------
\subsection{Experimental Setup}\label{subsec:exp_setup}
%-------------------------------
\textbf{Datasets and Models.} We conduct the experiments on two datasets: CIFAR-10~\cite{krizhevsky2009learning} and FEMNIST~\cite{caldas2019leaf}. Specifically, CIFAR-10 dataset consists of 60,000 color images classified into 10 categories. The dataset is divided into 50,000 training images and 10,000 testing images. FEMNIST, an extension of the MNIST dataset for federated learning, serves as a natural non-IID dataset with its 805,263 images unevenly distributed among 3,550 writers, each representing a client. We divide $90\%$ and $10\%$ of each client's data into training data and testing data, respectively. For CIFAR-10 dataset, we train a ResNet-20 model with 269,722 parameters in total. For FEMNIST, we adopt a CNN with 6,603,710 parameters in total~\cite{li2019abnormal}, which consists of two $3 \times 3$ convolutional layers (each with 32 channels and ReLu activation followed with $2 \times 2$ max pooling), a fully connected layer with 1024 units and ReLu activation, and a final softmax output layer.

\textbf{System Setting.}
We consider a CFEL system with 64 edge devices evenly distributed into 8 clusters. Edge servers are connected with a ring topology. For all experiments, each device updates its local model via mini-batch SGD with momentum of 0.9 and batch size of 50. The learning rate is tuned by the grid search from $\{0.01, 0.05, 0.1\}$ for CIAFR-10 and from
$\{0.1, 0.06, 0.03, 0.01\}$ for FEMNIST, respectively. 

\textbf{Data and System Heterogeneities.}
To simulate data heterogeneity, we partition the training images of CIFAR-10 to 64 devices by the Dirichlet distribution~\cite{hsu2019measuring} with a concentration parameter of $\beta$ (1.0 by default). FEMNIST has a natural non-IID partitioning of the data. Therefore, we randomly sample 64 devices to simulate the non-IID data distribution in our experiments. 

The total training time is calculated by combining the time spent on computing and the time required for communication. For measuring the computational workload, we utilize thop (available at https://pypi.org/project/thop/), which evaluates the workload in terms of floating point operations (FLOPs). For each training sample per iteration, the number of FLOPs needed for each training sample per iteration is 123.9 MFLOPs for ResNet-20 on CIFAR-10 dataset and 13.30 MFLOPs for training a CNN on the FEMNIST dataset. We base our calculations on the processing capabilities of the iPhone X, which has a processing capacity of 691.2 billion FLOPs (GFLOPs). We set the available CPU frequency to be randomly changed between \SI{1.0}{GHz} and \SI{2.0}{GHz} and the channel gain following the exponential distribution with a mean value of 1.0. The resulting computing time $\mu_n^{l,r}$ is between \SI{75}{s} and \SI{150}{s} and computing energy $\alpha_n^{l,r}$ is between \SI{1.5}{J} and \SI{6.0}{J}. Considering heterogeneous communication capabilities, we fluctuate each device's communication power $p_{n}$ between \SI{0.1}{W} and \SI{1.0}{W}. The white noise power spectral density
is \SI{0.01}{W}. To reflect heterogeneous devices and network conditions, we fluctuate each device's bandwidth between \SI{1}{Mbs} and \SI{5}{Mbs}~\cite{wang2022accelerating}. Following~\cite{sun2022semi}, we assume the bandwidth of backhaul between edge servers is \SI{50}{Mbs}. To achieve efficient computation and communication in the CFEL framework, we establish default time and energy budgets, denoted as $\tilde{\mathcal{T}}$ and $\tilde{\mathcal{E}}$, at 60\% of the benchmarks established by the SOTA algorithm in the CFEL framework, as referenced in~\cite{zhang2022scalable}. Specifically, \SI{8.5e4}{\second}
 and \SI{15}{KJ} for CIFAR-10 and \SI{1.3e5}{\second}
 and \SI{230}{KJ} for FEMNIST. Note that the budgets are inherently flexible and can be dynamically adjusted to better align with specific requirements and real-world constraints.

\textbf{Benchmarks.}
We compare the proposed framework with the following benchmarks.
\begin{itemize}
    \item \emph{CE-FedAvg (CEF)~\cite{zhang2022scalable}:} This is the state-of-the-art learning algorithm in CFEL that is oblivious to the system heterogeneity. Neither local update frequency control nor communication compression is considered. 
    \item \emph{CE-FedAvg with Adaptive Local Update Frequency Only (CEF-F):} In each edge round, the coordinator only controls local update probability to optimize the same problem as HCEF while transmitting the full model. This corresponds to HCEF when $\theta_n^{l,r}=1, \forall n, l, r$.
    \item \emph{CE-FedAvg with Adaptive Compression Only (CEF-C):} In each edge round, the coordinator only controls the local compression ratio to optimize the same problem as HCEF while performing the same steps of local SGD update for each device. This corresponds to HCEF when $\rho_n^{l,r}=1, \forall n, l, r$.
    \item \emph{MLL-SGD~\cite{castiglia2021multi}:} It is proposed to address resource heterogeneity by assigning each worker a probability of executing local SGD based on its computational capacity. This approach ensures that slower workers do not hinder the overall execution of the algorithm. When adapting to our setting, in each edge round, the coordinator assigns the local update probability inversely proportional to its computing time (i.e., $\rho_n^{l,r}=(1/\alpha_n^{l,r})/\sum_{n\in[N]}(1/\alpha_n^{l,r}$)).
\end{itemize}

%---------------------
\subsection{Experimental Results}\label{subsec_exp_results}
%---------------------

%%%%%%%%%%%%%%%%%%%%%%%%%%%%
\begin{figure}[t]
\subfloat[CIFAR-10]{{\includegraphics[width=0.22\textwidth]{ {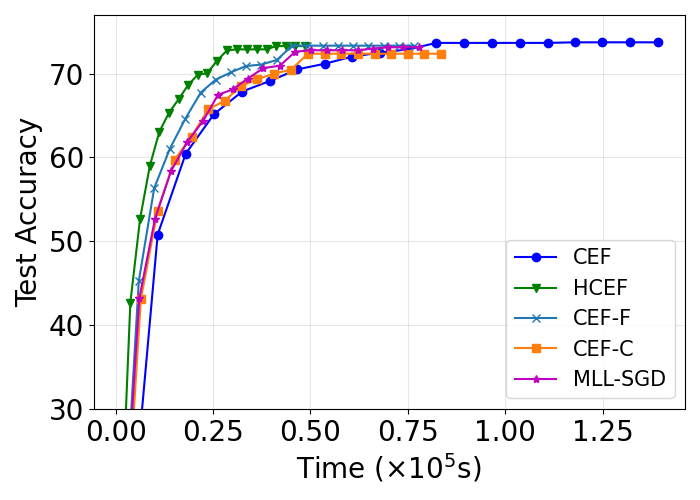}} }\label{fig:cifar_time}}
\subfloat[CIFAR-10]{{\includegraphics[width=0.22\textwidth]{ {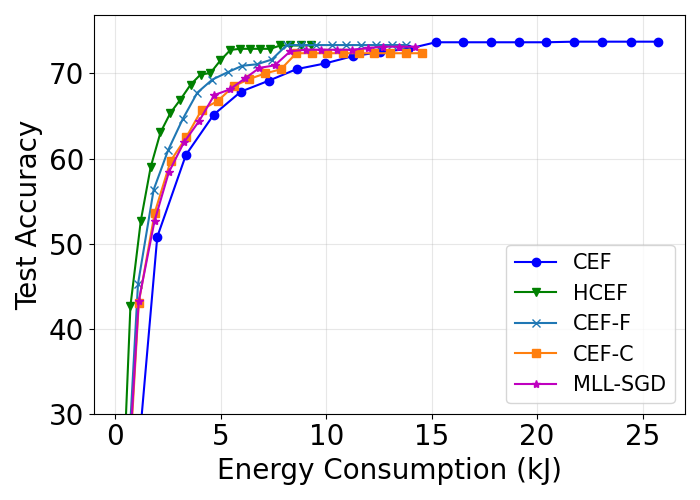}} }\label{fig:cifar_energy}}
\caption{Test accuracy versus runtime and energy consumption for CIFAR-10.
}\label{fig:cifar}
\vspace{-0.5cm}
\end{figure}

%%%%%%%%%%%%%%%%%%%%%%%%%%%%

\begin{figure}[t]
\subfloat[FEMNIST]{{\includegraphics[width=0.22\textwidth]{ {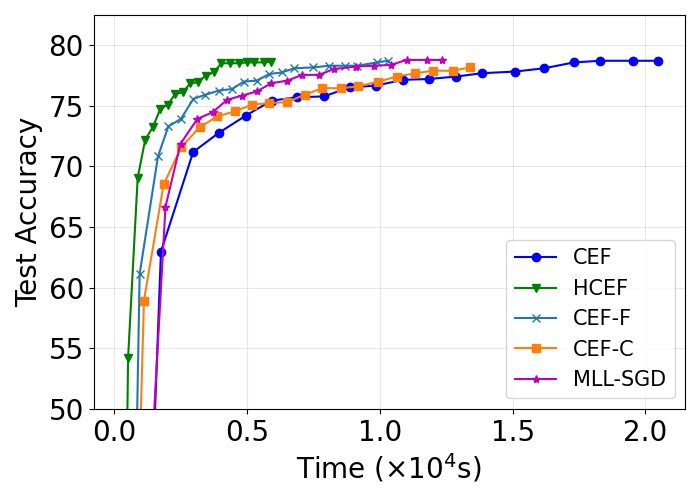}} }\label{fig:femnist_time}}
\subfloat[FEMNIST]{{\includegraphics[width=0.22\textwidth]{ {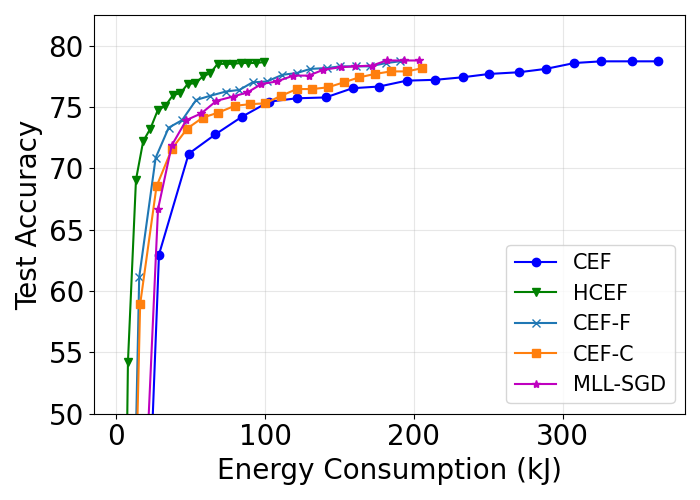}} }\label{fig:femnist_energy}}
\caption{Test accuracy versus runtime and energy consumption for FEMNIST.
}\label{fig:femnist}
\vspace{-0.5cm}
\end{figure}
%%%%%%%%%%%%%%%%%%%%%%%%%%

\textbf{Training Performance.} We first compare the runtime and energy consumption of HCEF and the baselines while fixing $\tau = 5$ and $q = 5$. For all schemes, we measure the average test accuracy of device models at each global round. Fig.~\ref{fig:cifar} and Fig.~\ref{fig:femnist} show the convergence processes on CIFAR-10 and FEMNIST, respectively. Specifically, Fig.~\ref{fig:cifar_time} illustrates that HCEF converges much faster than other baselines while maintaining a similar final accuracy with the CEF. Compared to CEF, we observe that HCEF is more energy-efficient in Fig.~\ref{fig:cifar_energy}. Note that CEF-F and MLL-SGD are also more time-efficient and energy-efficient than CEF. This is because they have reduced resource consumption in the computing phase. Similarly, CEF-C is also more time-efficient and energy-efficient than CEF due to reducing the resource consumption in the communication phase. Since MLL-SGD does not directly optimize accuracy, it is less effective compared to CEF-F. Typically, resource consumption is larger in the local computing stage than communication stage. By prioritizing resource reduction in the local computing phase, CEF-F and MLL-SGD are able to achieve better time efficiency and energy efficiency compared to CEF-C. CEF-C uses the same fixed local update frequency as CEF but compresses its model updates, which leads to a decrease in accuracy compared to CEF. Moreover, by jointly optimizing local update frequencies and compression ratios, HCEF consistently outperforms CEF-F, CEF-C, and MLL-SGD in terms of time efficiency and energy efficiency. For the FEMNIST dataset, we can observe similar results in Fig.~\ref{fig:femnist}.

\begin{table}[t]
\caption{Resource overhead of different methods to achieve the target accuracy.}
\begin{adjustbox}{width=\columnwidth,center}
\begin{tabular}{|c|c|c|c|c|c|c|}
\hline
Datasets & Metrics & CEF & CEF-C & CEF-F & MLL-SGD & \textbf{HCEF}\\
\hline
CIFAR-10 & Times($\times10^4$s)& 4.3 & 4.2 & 3.0 & 3.3 & \textbf{2.2(1.9$\times$)}\\
\cline{2-7}  
(Acc = 70\%) & Energy($\si{kJ}$) & 7.9 & 7.5 & 5.3 & 5.7 & \textbf{4.3(1.8$\times$)} \\
\hline
FEMNIST & Times($\times10^4$s)& 5.4 & 4.9 & 3.4 & 4.1 & \textbf{1.9(2.8$\times$})\\
\cline{2-7}  
(Acc = 75\%) & Energy($\si{kJ}$) & 93.5 & 73.6 & 48.0 & 65.2 & \textbf{30.3(1.9$\times$)} \\
\hline
\end{tabular}
\label{tab1}
\end{adjustbox}
\end{table}

\textbf{Resource Overhead.} To further validate the resource efficiency of HCEF, we record the time and energy consumption of HCEF and other methods when they meet the target accuracy in Table~\ref{tab1}. For the CIFAR-10 dataset, HCEF accelerates the training time by 1.9 times and reduces energy consumption by 1.8 times while achieving 70\% accuracy compared to CEF. This is due to the fact that HCEF uses adaptive local update frequencies and compression ratios to balance the trade-off between accuracy and resource consumption. CEF-F and MLL-SGD also reduce the time and energy consumption compared to CEF by adjusting the local update frequency. Typically, resource consumption is higher in the local computing stage than in the communication stage. By prioritizing resource reduction during local computing, CEF-F and MLL-SGD achieve better time and energy efficiency compared to CEF-C, which reduces the resource consumption in the communication phase. In summary, HCEF brings significant savings in both time and energy consumption while maintaining high accuracy. Similar results can be verified in the FEMNIST dataset.

\begin{figure}[t]
\subfloat[Runtime]{{\includegraphics[width=0.22\textwidth]{ {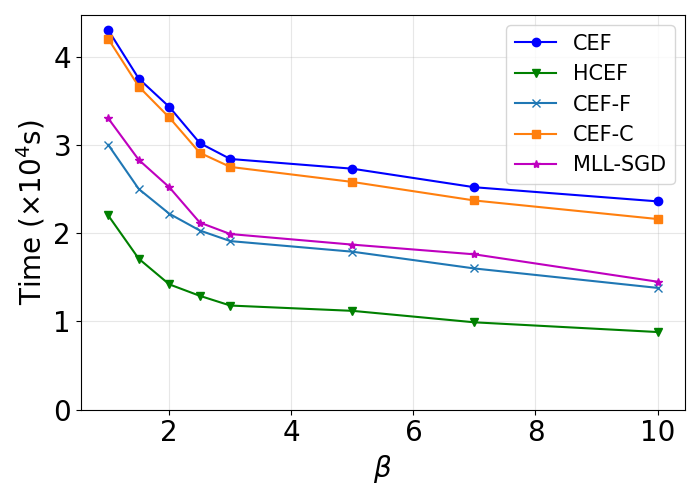}} }\label{fig:cifar_time_noniid}}
\subfloat[Energy]{{\includegraphics[width=0.22\textwidth]{ {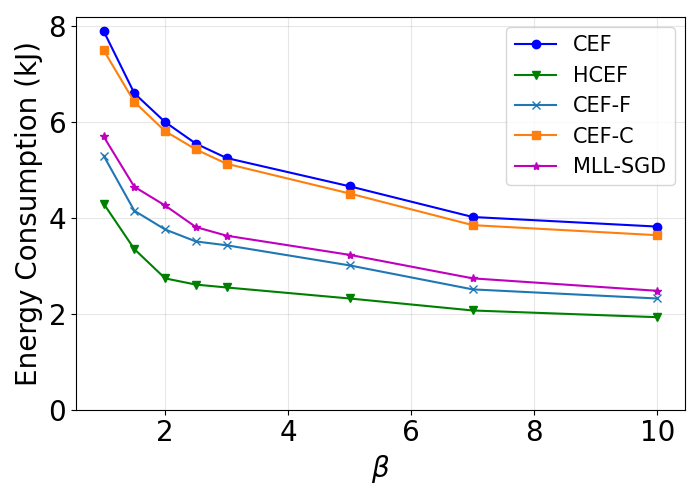}} }\label{fig:cifar_energy_noniid}}
\caption{Runtime and energy consumption under different non-IID levels for CIFAR-10 with target accuracy $70\%$.
}\label{fig:cifar_noniid}
\vspace{-0.6cm}
\end{figure}

\textbf{Effect of Non-IID Data.} We further investigate the performance of our algorithm under statistical heterogeneity. To examine the effect of non-IID data distribution, we conduct experiments on CIFAR-10 by varying the concentration parameter $\beta$. Smaller values of $\beta$ result in higher levels of data heterogeneity. Fig.~\ref{fig:cifar_noniid} depicts the required time and energy for HCEF and other baselines to achieve the target accuracy of 70\%. As shown in Fig.~\ref{fig:cifar_noniid}, all methods experience larger time and energy consumption as the data distribution's skewness increases (smaller $\beta$). However, HCEF always achieves significant savings in both training time and energy consumption across different statistical heterogeneity levels. In addition, the savings of resource overhead (either time or energy) in HCEF further enlarge as the non-IID level increases.

%---------------------
\subsection{Ablation Studies}\label{sub_sec_ablation}
\begin{figure}[t]
\subfloat[CIFAR-10]{{\includegraphics[width=0.22\textwidth]{ {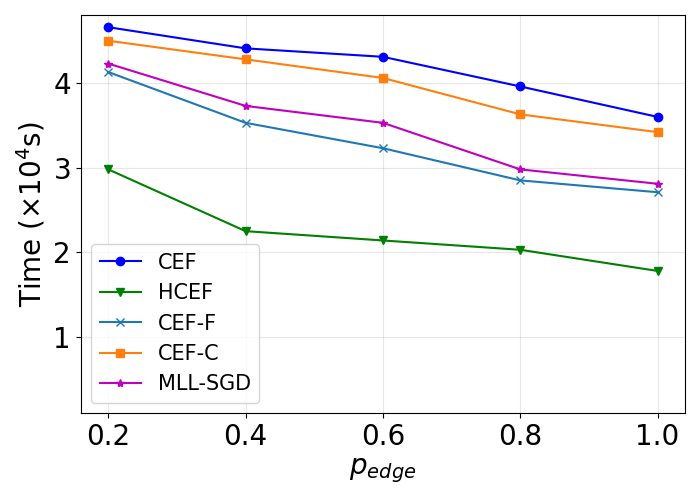}} }\label{fig:cifar_time_topo}}
\subfloat[CIFAR-10]{{\includegraphics[width=0.22\textwidth]{ {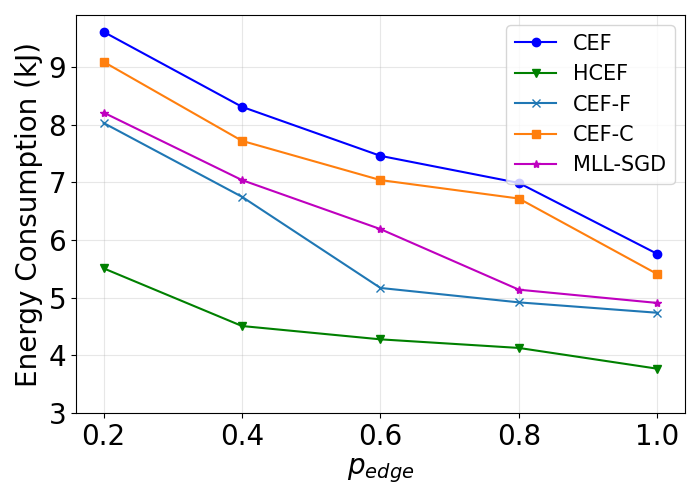}} }\label{fig:cifar_energy_topo}}\\
[-2.5ex]
\subfloat[FEMNIST]{{\includegraphics[width=0.22\textwidth]{ {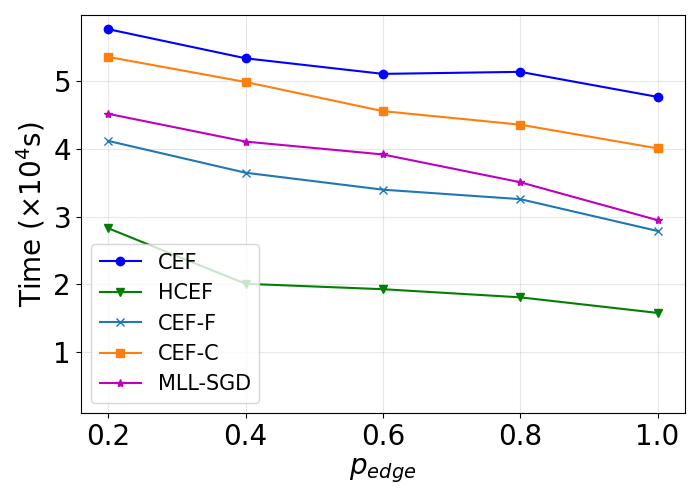}} }\label{fig:femnist_time_topo}}
\subfloat[FEMNIST]{{\includegraphics[width=0.22\textwidth]{ {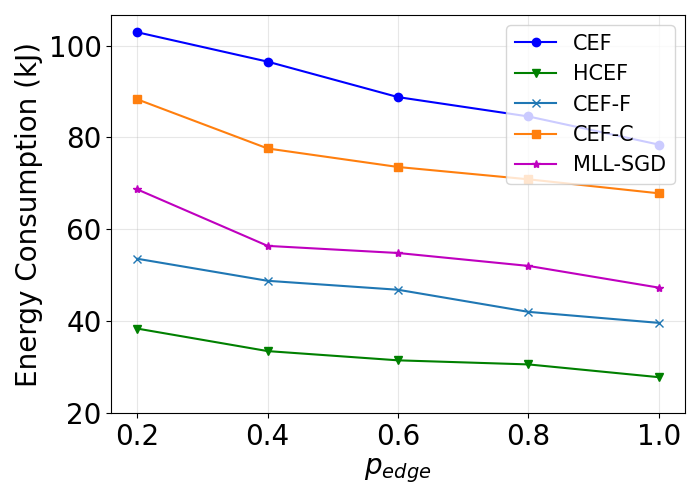}} }\label{fig:femnist_energy_topo}}
\caption{(a), (b) Runtime and energy consumption under different backhaul topologies for CIFAR-10 with target accuracy $70\%$; (c), (d) Runtime and energy consumption under different backhaul topologies for FEMNIST with target accuracy $75\%$.
}\label{fig:topology}
\vspace{-0.6cm} 
\end{figure}

%---------------------
\textbf{Effect of Backhaul topology.}
We further evaluate the performance of our algorithm and baselines under varying edge backhaul topologies in Fig.~\ref{fig:topology}. We generate random network topologies by Erd\H{o}s-R\'enyi model with edge probability $p_{edge}=\{0.2, 0.4, 0.6,0.8,1.0\}$. Higher values of $p_{edge}$ result in increased graph connectivity. Notably, at $p_{edge}=1.0$, the graph becomes fully connected. Enhanced connectivity facilitates more efficient data sharing and coordination among devices, which in turn reduces the training rounds of our algorithm. Consequently, higher values $p_{edge}$ lead to substantial savings in both total training time and energy usage. As observed in the figure, a more connected backhaul topology (i.e., a larger value of $p_{edge}$) generally accelerates the convergence and achieves significant savings in both training time and energy consumption. In addition, HCEF always achieves significant savings in both training time and energy consumption across different backhaul topologies.

%%%%%%%%%%%%%%%%%%%%%%%%%%%%
%%%%%%%%%%%%%%%%%%%%%%%%%%%%
%%%%%%%%%%%%%%%%%%%%%%%%%%%%
%%%%%%%%%%%%%%%%%%%%%%%%%%%%

\begin{figure}[t]
\subfloat[CIFAR-10]{{\includegraphics[width=0.22\textwidth]{ {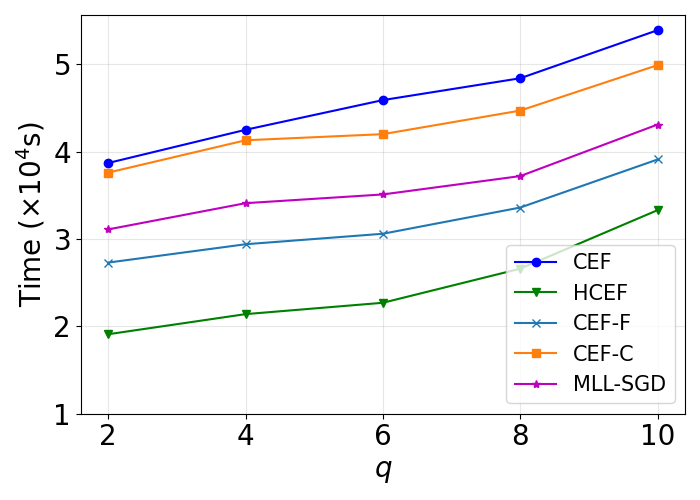}} }\label{fig:cifar_time_q}}
\subfloat[CIFAR-10]{{\includegraphics[width=0.22\textwidth]{ {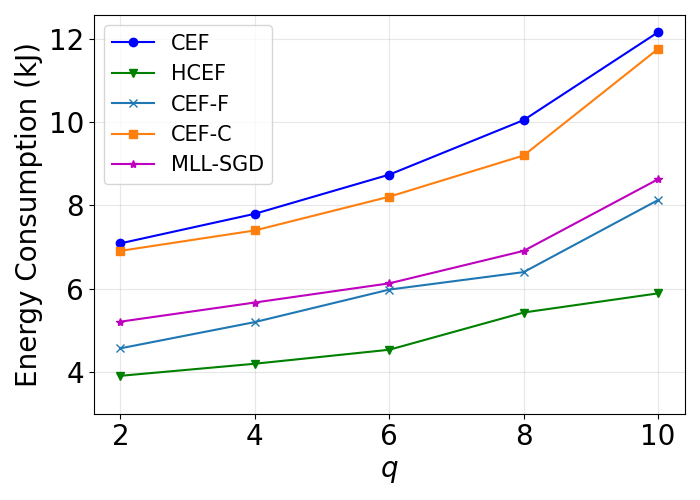}} }\label{fig:cifar_energy_q}}\\
[-2.5ex]
\subfloat[FEMNIST]{{\includegraphics[width=0.22\textwidth]{ {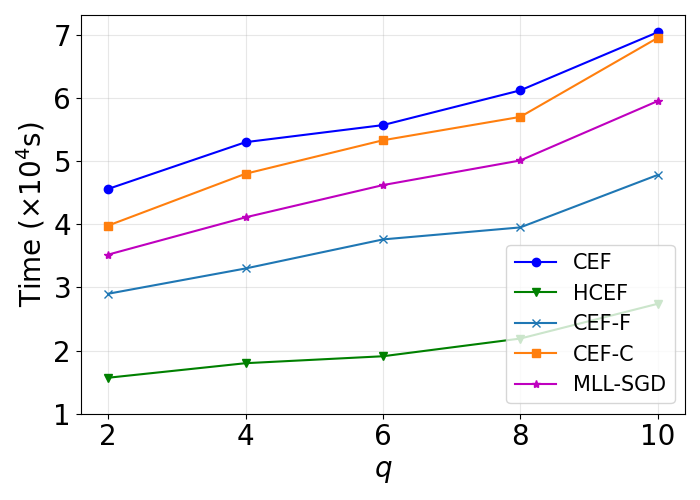}} }\label{fig:femnist_time_q}}
\subfloat[FEMNIST]{{\includegraphics[width=0.22\textwidth]{ {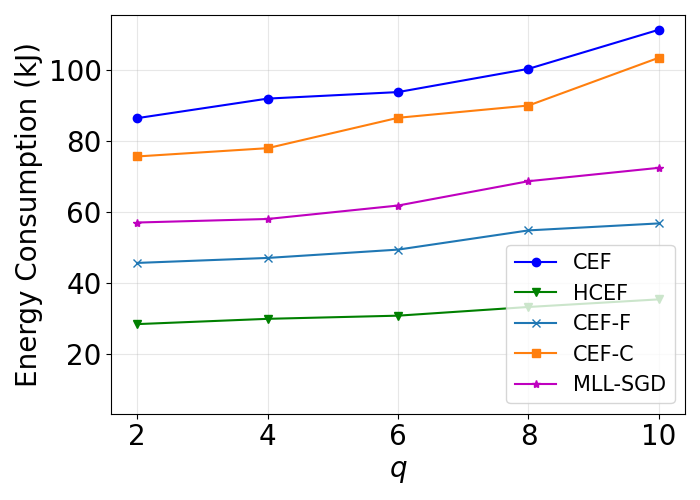}} }\label{fig:femnist_energy_q}}
\caption{(a), (b) Runtime and energy consumption under different $q$ for CIFAR-10 with target accuracy $70\%$; (c), (d) Runtime and energy consumption under different $q$ for FEMNIST with target accuracy $75\%$.
}\label{fig:intra_peroid}
\vspace{-0.6cm} 
\end{figure}

\begin{figure}[t]
\subfloat[CIFAR-10]{{\includegraphics[width=0.22\textwidth]{ {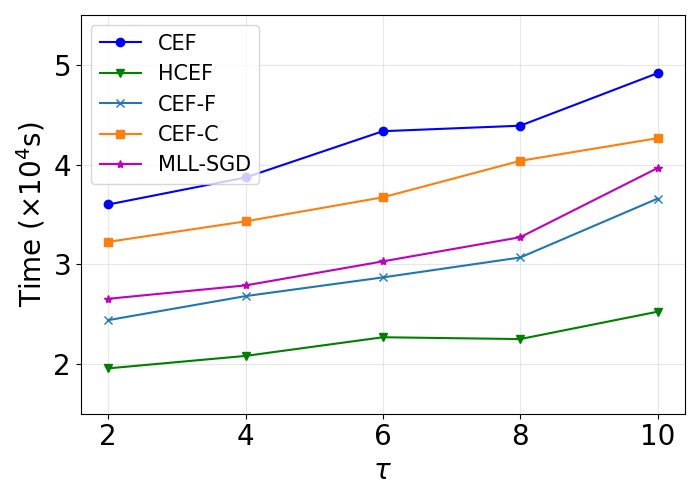}} }\label{fig:cifar_time_tau}}
\subfloat[CIFAR-10]{{\includegraphics[width=0.22\textwidth]{ {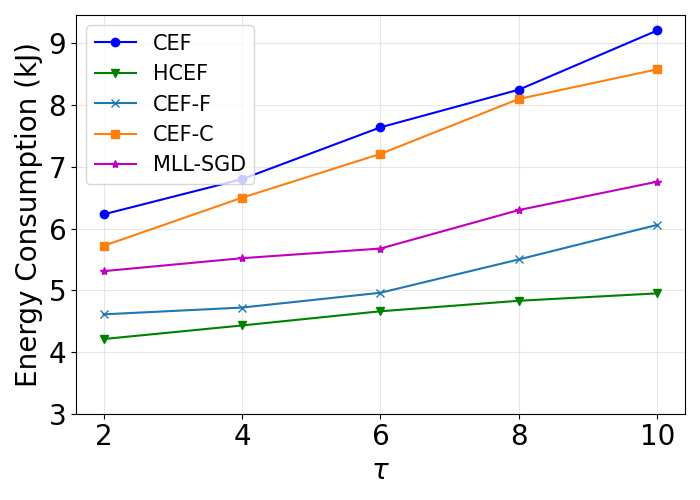}} }\label{fig:cifar_energy_tau}}\\
[-2.5ex]
\subfloat[FEMNIST]{{\includegraphics[width=0.22\textwidth]{ {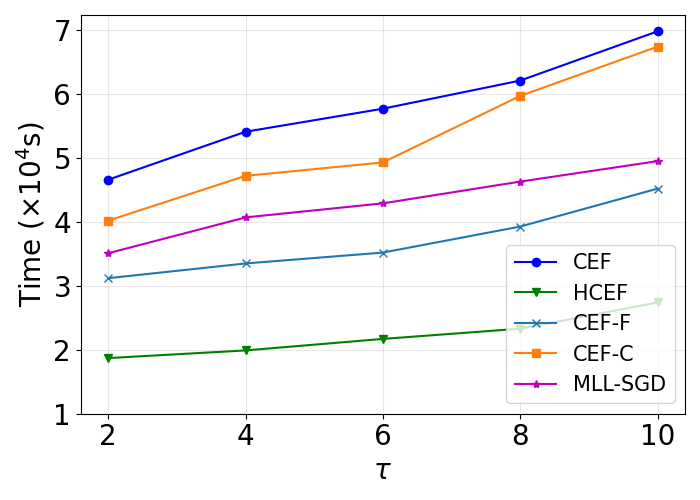}} }\label{fig:femnist_time_tau}}
\subfloat[FEMNIST]{{\includegraphics[width=0.22\textwidth]{ {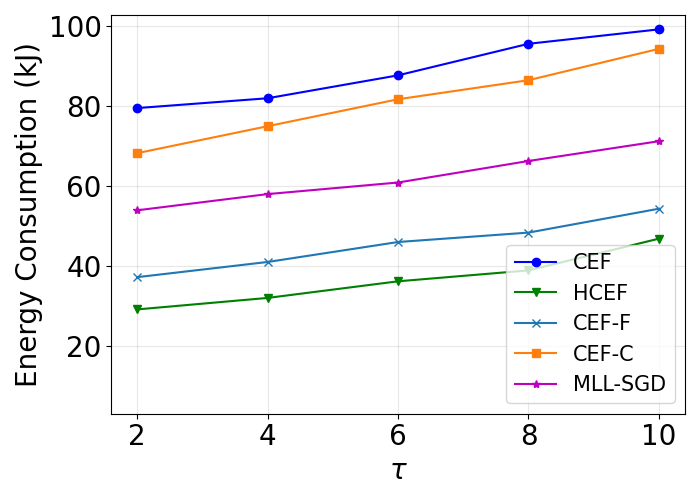}} }\label{fig:femnist_energy_tau}}
\caption{(a), (b) Runtime and energy consumption under different $\tau$ for CIFAR-10 with target accuracy $70\%$; (c), (d) Runtime and energy consumption under different $\tau$ for FEMNIST with target accuracy $75\%$.
}\label{fig:inter_peroid}
\vspace{-0.6cm}   
\end{figure}

\textbf{Effect of aggregation period.}
We further evaluate the performance of our algorithm and baseline methods across different intra-cluster and inter-cluster aggregation periods, as illustrated in Fig.~\ref{fig:intra_peroid},~\ref{fig:inter_peroid}. We have set different intra-cluster aggregation period $\tau = \{2, 4, 6, 8, 10\}$ with a default $q=5$ and the inter-cluster aggregation period by $q = \{2, 4, 6, 8, 10\}$ with $\tau=5$. Higher values of $q, \tau$ result in decreased aggregation periods. Reduced aggregation period inhibits efficient data sharing and coordination among devices, which in turn increases the training rounds of our algorithm. This can also lead to reduced communication time and energy consumption. Therefore, when the increase in computing time and energy is offset by the reduction in communication time and energy, higher values of $q, \tau$ ultimately lead to increased total training time and energy usage. As observed in the figures, higher values of $q, \tau$ generally accelerate the training time and energy consumption. In addition, HCEF always achieves significant savings in both training time and energy consumption across different $q, \tau$.

%%%%%%%%%%%%%%%%%%%%%%%%%%%%

%%%%%%%%%%%%%%%%%%%%%%%%%%%%

%%%%%%%%%%%%%%%%%%%%%%%%%%%%
%%%%%%%%%%%%%%%%%%%%%%%%%%%%
{\textbf{Training parameters $\sigma^2$ and $G^2$.} Fig.~\ref{fig:sig_g} illustrates the estimation of the training parameters $\sigma^2$ and $G^2$ during the execution of the proposed HCEF algorithm. $\sigma^2$ represents the variance of the local mini-batch stochastic gradient, while $G^2$ tracks the gradient magnitude across. The decreasing trend of $G^2$ indicates that our algorithm converges efficiently.
%%%%%%%%%%%%%%%%%%%%%%%%%%%%
\begin{figure}[t]
\centering
\subfloat[CIFAR-10]{{\includegraphics[width=0.22\textwidth]{ {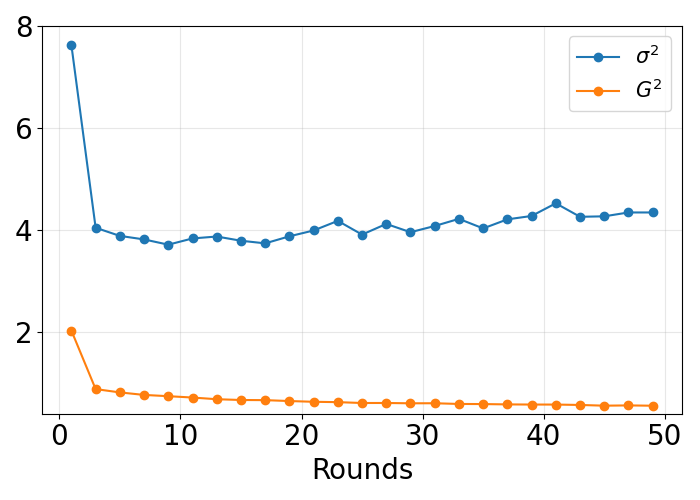}} }\label{fig:cifar_sigma}}
\subfloat[FEMNIST]{{\includegraphics[width=0.22\textwidth]{ {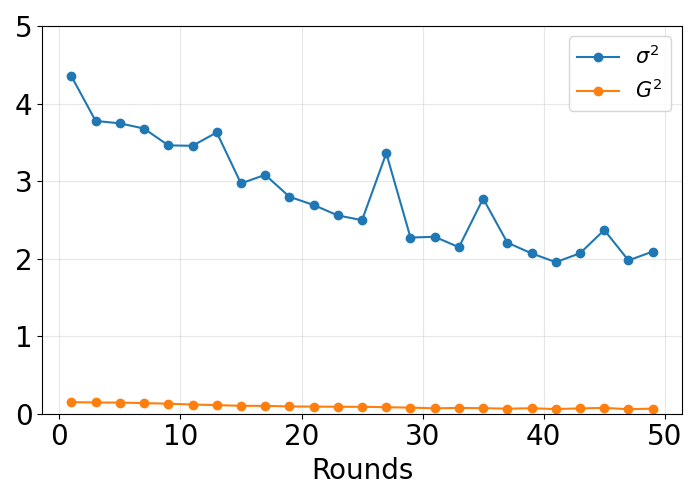}} }\label{fig:femnist_sigma}}
\caption{Training parameters $\sigma^2$ and $G^2$ versus training round.
}\label{fig:sig_g}
\end{figure}
%%%%%%%%%%%%%%%%%%%%%%%%%%%%
%%%%%%%%%%%%%%%%%%%%%%%%%%%%%%%%%
\section{Conclusions}\label{sec:conclusion}
%%%%%%%%%%%%%%%%%%%%%%%%%%%%%%%%%
In this paper, we have proposed an efficient scheme named HCEF that integrates adaptive control of local update frequency and gradient compression to effectively address the challenges posed by system heterogeneity and dynamic state in CFEL. By capturing the trade-off between accuracy and resource consumption, we have developed an efficient online control algorithm to dynamically determine local update frequencies and compression ratios. We have evaluated our method through extensive experiments based on common FL benchmark datasets and demonstrated that HCEF can learn an accurate model within a shorter time and lower energy consumption than other FL frameworks over mobile edge networks.

%%%%%%%%%%%%%%%%%%%%%%%%%%%%
\section{ACKNOWLEDGMENTS}
%%%%%%%%%%%%%%%%%%%%%%%%%%%%
The work of Z. Zhang, Z. Gao, and Y. Gong was partially supported by the US National Science Foundation under Grant CNS-2047761 and Grant CNS-2106761. The work of Y. Guo was partially supported by the US National Science Foundation under Grant CNS-CNS-2106761 and CNS-2318683.

% or



% % you can choose not to have a title for an appendix
% % if you want by leaving the argument blank
% \section{}

% % use section* for acknowledgment
% \ifCLASSOPTIONcompsoc
%   % The Computer Society usually uses the plural form
%   \section*{Acknowledgments}
% \else
%   % regular IEEE prefers the singular form
% \fi


\ifCLASSOPTIONcaptionsoff
  \newpage
\fi



%\bibliographystyle{IEEEtran}
\bibliographystyle{IEEEtran}
\bibliography{main.bib}

\begin{thebibliography}{10}
\providecommand{\url}[1]{#1}
\csname url@samestyle\endcsname
\providecommand{\newblock}{\relax}
\providecommand{\bibinfo}[2]{#2}
\providecommand{\BIBentrySTDinterwordspacing}{\spaceskip=0pt\relax}
\providecommand{\BIBentryALTinterwordstretchfactor}{4}
\providecommand{\BIBentryALTinterwordspacing}{\spaceskip=\fontdimen2\font plus
\BIBentryALTinterwordstretchfactor\fontdimen3\font minus \fontdimen4\font\relax}
\providecommand{\BIBforeignlanguage}[2]{{%
\expandafter\ifx\csname l@#1\endcsname\relax
\typeout{** WARNING: IEEEtran.bst: No hyphenation pattern has been}%
\typeout{** loaded for the language `#1'. Using the pattern for}%
\typeout{** the default language instead.}%
\else
\language=\csname l@#1\endcsname
\fi
#2}}
\providecommand{\BIBdecl}{\relax}
\BIBdecl

\bibitem{mcmahan2017communication}
B.~McMahan, E.~Moore, D.~Ramage, S.~Hampson, and B.~A. y~Arcas, ``Communication-efficient learning of deep networks from decentralized data,'' in \emph{Proc. Int. Artif. Intell. Statist.}, 2017, pp. 1273--1282.

\bibitem{zhang2024energy}
Y.~Zhang, Y.~Gong, and Y.~Guo, ``Energy-efficient resource management for multi-{UAV}-enabled mobile edge computing,'' \emph{IEEE Trans. Veh. Technol.}, vol.~73, no.~8, pp. 12\,026--12\,037, 2024.

\bibitem{yu2020joint}
Z.~Yu, Y.~Gong, S.~Gong, and Y.~Guo, ``Joint task offloading and resource allocation in {UAV}-enabled mobile edge computing,'' \emph{IEEE Internet Things J.}, vol.~7, no.~4, pp. 3147--3159, 2020.

\bibitem{ding2018beef}
H.~Ding, Y.~Guo, X.~Li, and Y.~Fang, ``Beef up the edge: Spectrum-aware placement of edge computing services for the internet of things,'' \emph{IEEE Trans. Mobile Comput.}, vol.~18, no.~12, pp. 2783--2795, 2018.

\bibitem{kairouz2021advances}
P.~Kairouz, H.~B. McMahan, B.~Avent, A.~Bellet, M.~Bennis, A.~N. Bhagoji, K.~Bonawitz, Z.~Charles, G.~Cormode, R.~Cummings \emph{et~al.}, ``Advances and open problems in federated learning,'' \emph{Foundations and Trends{\textregistered} in Machine Learning}, vol.~14, no. 1--2, pp. 1--210, 2021.

\bibitem{castiglia2021multi}
T.~Castiglia, A.~Das, and S.~Patterson, ``Multi-level local {SGD}: Distributed {SGD} for heterogeneous hierarchical networks,'' in \emph{Proc. Int. Conf. Learn. Represent. (ICLR)}, 2021.

\bibitem{sun2023semi}
Y.~Sun, J.~Shao, Y.~Mao, J.~H. Wang, and J.~Zhang, ``Semi-decentralized federated edge learning with data and device heterogeneity,'' \emph{IEEE Trans. Netw. Service Manag.}, vol.~20, no.~2, pp. 1487--1501, 2023.

\bibitem{zhong2021p}
Z.~Zhong, Y.~Zhou, D.~Wu, X.~Chen, M.~Chen, C.~Li, and Q.~Z. Sheng, ``P-{F}ed{A}vg: {P}arallelizing federated learning with theoretical guarantees,'' in \emph{Proc. IEEE Conf. Comput. Commun. (INFOCOM)}, 2021, pp. 1--10.

\bibitem{zhang2022scalable}
Z.~Zhang, Z.~Gao, Y.~Guo, and Y.~Gong, ``Scalable and low-latency federated learning with cooperative mobile edge networking,'' \emph{IEEE Trans. Mobile Comput.}, vol.~23, no.~1, pp. 812--822, 2022.

\bibitem{nori2021fast}
M.~K. Nori, S.~Yun, and I.-M. Kim, ``Fast federated learning by balancing communication trade-offs,'' \emph{IEEE Trans. Commun.}, vol.~69, no.~8, pp. 5168--5182, 2021.

\bibitem{xu2022adaptive}
Y.~Xu, Y.~Liao, H.~Xu, Z.~Ma, L.~Wang, and J.~Liu, ``Adaptive control of local updating and model compression for efficient federated learning,'' \emph{IEEE Trans. Mobile Comput.}, vol.~22, no.~10, pp. 5675--5689, 2022.

\bibitem{wang2018atomo}
H.~Wang, S.~Sievert, S.~Liu, Z.~Charles, D.~Papailiopoulos, and S.~Wright, ``{ATOMO}: Communication-efficient learning via atomic sparsification,'' in \emph{Proc. Adv. Neural Inf. Process. Syst.}, 2018, pp. 9850--9861.

\bibitem{sonee2021wireless}
A.~Sonee, S.~Rini, and Y.-C. Huang, ``Wireless federated learning with limited communication and differential privacy,'' in \emph{Proc. IEEE Glob. Commun. Conf. (GLOBECOM)}, 2021, pp. 01--06.

\bibitem{shlezinger2020uveqfed}
N.~Shlezinger, M.~Chen, Y.~C. Eldar, H.~V. Poor, and S.~Cui, ``{UVeQFed}: Universal vector quantization for federated learning,'' \emph{IEEE Trans. Signal Process.}, vol.~69, pp. 500--514, 2020.

\bibitem{reisizadeh2020fedpaq}
A.~Reisizadeh, A.~Mokhtari, H.~Hassani, A.~Jadbabaie, and R.~Pedarsani, ``{FedPAQ}: A communication-efficient federated learning method with periodic averaging and quantization,'' in \emph{Proc. Int. Conf. Artif. Intell. Statist.}, 2020, pp. 2021--2031.

\bibitem{han2020adaptivegrad}
P.~Han, S.~Wang, and K.~K. Leung, ``Adaptive gradient sparsification for efficient federated learning: An online learning approach,'' in \emph{Proc. IEEE 40th Int. Conf. Distrib. Comput. Syst. (ICDCS)}.\hskip 1em plus 0.5em minus 0.4em\relax IEEE, 2020, pp. 300--310.

\bibitem{cui2022optimal}
L.~Cui, X.~Su, Y.~Zhou, and J.~Liu, ``Optimal rate adaption in federated learning with compressed communications,'' in \emph{Proc. IEEE Conf. Comput. Commun. (INFOCOM)}.\hskip 1em plus 0.5em minus 0.4em\relax IEEE, 2022, pp. 1459--1468.

\bibitem{li2021talk}
L.~Li, D.~Shi, R.~Hou, H.~Li, M.~Pan, and Z.~Han, ``To talk or to work: Flexible communication compression for energy efficient federated learning over heterogeneous mobile edge devices,'' in \emph{Proc. IEEE Conf. Comput. Commun. (INFOCOM)}.\hskip 1em plus 0.5em minus 0.4em\relax IEEE, 2021, pp. 1--10.

\bibitem{zhang2023communication}
Z.~Zhang, Y.~Guo, Y.~Fang, and Y.~Gong, ``Communication and energy efficient wireless federated learning with intrinsic privacy,'' \emph{IEEE Trans. Depend. Sec. Comput.}, 2023.

\bibitem{luo2021cost}
B.~Luo, X.~Li, S.~Wang, J.~Huang, and L.~Tassiulas, ``Cost-effective federated learning design,'' in \emph{Proc. IEEE Conf. Computing. Commun. (INFOCOM)}.\hskip 1em plus 0.5em minus 0.4em\relax IEEE, 2021, pp. 1--10.

\bibitem{wang2019adaptive}
S.~Wang, T.~Tuor, T.~Salonidis, K.~K. Leung, C.~Makaya, T.~He, and K.~Chan, ``Adaptive federated learning in resource constrained edge computing systems,'' \emph{IEEE J. Sel. Areas Commun.}, vol.~37, no.~6, pp. 1205--1221, 2019.

\bibitem{wang2019matcha}
J.~Wang, A.~K. Sahu, Z.~Yang, G.~Joshi, and S.~Kar, ``{MATCHA}: Speeding up decentralized {SGD} via matching decomposition sampling,'' in \emph{Proc. 6th Indian Control Conf.}\hskip 1em plus 0.5em minus 0.4em\relax IEEE, 2019, pp. 299--300.

\bibitem{wang2022accelerating}
L.~Wang, Y.~Xu, H.~Xu, M.~Chen, and L.~Huang, ``Accelerating decentralized federated learning in heterogeneous edge computing,'' \emph{IEEE Trans. Mobile Comput.}, vol.~22, no.~9, pp. 5001--5016, 2022.

\bibitem{xu2021decentralized}
H.~Xu, M.~Chen, Z.~Meng, Y.~Xu, L.~Wang, and C.~Qiao, ``Decentralized machine learning through experience-driven method in edge networks,'' \emph{IEEE J. Sel. Areas Commun.}, vol.~40, no.~2, pp. 515--531, 2021.

\bibitem{lyu2018multi}
X.~Lyu, C.~Ren, W.~Ni, H.~Tian, R.~P. Liu, and Y.~J. Guo, ``Multi-timescale decentralized online orchestration of software-defined networks,'' \emph{IEEE J. Sel. Areas Commun.}, vol.~36, no.~12, pp. 2716--2730, 2018.

\bibitem{stich2018sparsified}
S.~U. Stich, J.-B. Cordonnier, and M.~Jaggi, ``Sparsified {SGD} with memory,'' in \emph{Proc. Adv. Neural Inf. Process. Syst.}, 2018, pp. 4447--4458.

\bibitem{hu2023federated}
R.~Hu, Y.~Guo, and Y.~Gong, ``Federated learning with sparsified model perturbation: Improving accuracy under client-level differential privacy,'' \emph{IEEE Trans. Mobile Comput.}, vol.~23, no.~8, pp. 8242--8255, 2024.

\bibitem{m2021efficient}
A.~M~Abdelmoniem, A.~Elzanaty, M.-S. Alouini, and M.~Canini, ``An efficient statistical-based gradient compression technique for distributed training systems,'' \emph{Proc. Mach. Learn. Syst.}, vol.~3, pp. 297--322, 2021.

\bibitem{jiang2023heterogeneity}
Z.~Jiang, Y.~Xu, H.~Xu, Z.~Wang, and C.~Qian, ``Heterogeneity-aware federated learning with adaptive client selection and gradient compression,'' in \emph{Proc. IEEE Conf. Comput. Commun. (INFOCOM)}, 2023, pp. 1--10.

\bibitem{yang2020energy}
Z.~Yang, M.~Chen, W.~Saad, C.~S. Hong, and M.~Shikh-Bahaei, ``Energy efficient federated learning over wireless communication networks,'' \emph{IEEE Trans. Wireless Commun.}, vol.~20, no.~3, pp. 1935--1949, 2020.

\bibitem{nedic2018network}
A.~Nedi{\'c}, A.~Olshevsky, and M.~G. Rabbat, ``Network topology and communication-computation tradeoffs in decentralized optimization,'' \emph{Proc. IEEE}, vol. 106, no.~5, pp. 953--976, 2018.

\bibitem{bottou2018optimization}
L.~Bottou, F.~E. Curtis, and J.~Nocedal, ``Optimization methods for large-scale machine learning,'' \emph{Siam Review}, vol.~60, no.~2, pp. 223--311, 2018.

\bibitem{guo2022hybrid}
Y.~Guo, Y.~Sun, R.~Hu, and Y.~Gong, ``Hybrid local {SGD} for federated learning with heterogeneous communications,'' in \emph{Proc. Int. Conf. Learn. Represent. (ICLR)}, 2022.

\bibitem{koloskova2020unified}
A.~Koloskova, N.~Loizou, S.~Boreiri, M.~Jaggi, and S.~Stich, ``A unified theory of decentralized {SGD} with changing topology and local updates,'' in \emph{Proc. Int. Conf. Mach. Learn. (ICML)}, 2020, pp. 5381--5393.

\bibitem{wang2021cooperative}
J.~Wang and G.~Joshi, ``Cooperative {SGD}: A unified framework for the design and analysis of local-update {SGD} algorithms,'' \emph{J. Mach. Learn. Res.}, vol.~22, no. 213, pp. 1--50, 2021.

\bibitem{yu2019parallel}
H.~Yu, S.~Yang, and S.~Zhu, ``Parallel restarted {SGD} with faster convergence and less communication: Demystifying why model averaging works for deep learning,'' in \emph{Proc. Conf. Artif. Intell.}, vol.~33, no.~01, 2019, pp. 5693--5700.

\bibitem{tran2019federated}
N.~H. Tran, W.~Bao, A.~Zomaya, N.~M. NH, and C.~S. Hong, ``Federated learning over wireless networks: Optimization model design and analysis,'' in \emph{Proc. IEEE Conf. Comput. Commun. (INFOCOM)}.\hskip 1em plus 0.5em minus 0.4em\relax IEEE, 2019, pp. 1387--1395.

\bibitem{chen2020joint}
M.~Chen, Z.~Yang, W.~Saad, C.~Yin, H.~V. Poor, and S.~Cui, ``A joint learning and communications framework for federated learning over wireless networks,'' \emph{IEEE Trans. Wireless Commun.}, vol.~20, no.~1, pp. 269--283, 2020.

\bibitem{diamond2016cvxpy}
S.~Diamond and S.~Boyd, ``{CVXPY}: A python-embedded modeling language for convex optimization,'' \emph{J. Mach. Learn. Res.}, vol.~17, no.~1, pp. 2909--2913, 2016.

\bibitem{wang2014outage}
K.-Y. Wang, A.~M.-C. So, T.-H. Chang, W.-K. Ma, and C.-Y. Chi, ``Outage constrained robust transmit optimization for multiuser miso downlinks: Tractable approximations by conic optimization,'' \emph{IEEE Trans. Signal Process.}, vol.~62, no.~21, pp. 5690--5705, 2014.

\bibitem{krizhevsky2009learning}
A.~Krizhevsky, G.~Hinton \emph{et~al.}, ``Learning multiple layers of features from tiny images,'' 2009.

\bibitem{caldas2019leaf}
S.~Caldas, P.~Wu, T.~Li, J.~Kone{\v{c}}n{\`y}, H.~B. McMahan, V.~Smith, and A.~Talwalkar, ``{LEAF}: A benchmark for federated settings,'' in \emph{Proc. Workshop Federated Learn. Data Privacy Confidentiality}, 2019.

\bibitem{li2019abnormal}
S.~Li, Y.~Cheng, Y.~Liu, W.~Wang, and T.~Chen, ``Abnormal client behavior detection in federated learning,'' in \emph{Proc. Int. Workshop Federated Learn. Data Privacy Confidentiality}, 2019.

\bibitem{hsu2019measuring}
\BIBentryALTinterwordspacing
H.~Hsu, H.~Qi, and M.~Brown, ``Measuring the effects of non-identical data distribution for federated visual classification,'' 2019. [Online]. Available: \url{https://arxiv.org/abs/1909.06335}
\BIBentrySTDinterwordspacing

\bibitem{sun2022semi}
Y.~Sun, J.~Shao, Y.~Mao, J.~H. Wang, and J.~Zhang, ``Semi-decentralized federated edge learning for fast convergence on non-iid data,'' in \emph{Proc. IEEE Wireless Commun. Netw. Conf. (WCNC)}, 2022, pp. 1898--1903.

\end{thebibliography}

\newpage
\onecolumn
\title{Heterogeneity-Aware Cooperative Federated Edge Learning with Adaptive Computation and Communication Compression (Supplementary)}

% \markboth{}%
% {Shell \MakeLowercase{\textit{et al.}}: }

% \IEEEtitleabstractindextext{%
% }
% % \onecolumn
% % make the title area
% \maketitle

\appendices
%%%%%%%%%%%%%%%%%%%%%%%%%%%%%%%%%%%%%%%%%%%%%%%%%%%%%%%%%%%%%%%%%%%%%%%%%%%%%%%%%%%%%%%%%%%%%%%%%%%%%%%%%%%%%%%%%%%%%%%%%%%%%%%%%%%%%%%%%%%%%%%%%%%%%%%%%%%%%%%%%%%%%%%%%%%%%%%%%%%%%%%%%%%%%%%%%
%------------------------------------------------
\section{Update Rule for HCEF Scheme}
%------------------------------------------------

Since edge servers are essentially stateless in HCEF, we focus on how device models evolve in the convergence analysis. We define $t = l q\tau+r\tau+s$, where $ l \in [0, \phi-1]$, $r \in [0, q-1]$ and $s \in [0, \tau-1]$, as the global iteration index, and $T= pq\tau$ as the total number of global training iterations in Algorithm~\ref{algorithm-1}. Then we can rewritten the local model $\mathbf{x}_{n}^{l, r, s}$ as $\mathbf{x}_{n}^t$. Without loss of generality, we denote the range of device indices for cluster $i \in [m]$ as $\left[\sum_{j \leq i - 1} n_{j} + 1, \sum_{j \leq i} n_{j}\right]$ with $n_0 = 0$.

The system behavior of HCEF can be summarized by the following update rule for device models:
\setcounter{equation}{15}
\begin{equation}\label{eq_update_rule}
Q(\mathbf{X}^{t+1})= Q(\mathbf{X}^t-\eta\mathbf{G}^t)\mathbf{W}^t,
\end{equation}
where $Q=Q(\cdot, \theta_{n}^t; \forall n\in[N])$ is the compression operator when $(t+1)\mod \tau = 0$, $\mathbf{X}^t=[\mathbf{x}_1^t,\dots,\mathbf{x}_N^t]\in \mathbb{R}^{d\times N}$, $\mathbf{G}^t = [\mathbf{g}_1(\mathbf{x}_1^{t}),\dots,\mathbf{g}_N(\mathbf{x}_N^t)] \in \mathbb{R}^{d\times N}$, and $\mathbf{W}^t \in \mathbb{R}^{N\times N}$ is a time-varying operator capturing the three stages in HCEF: SGD update, intra-cluster model aggregation, and inter-cluster model aggregation. Specifically, $\mathbf{W}^t$ is defined as follows:
\begin{align}
\mathbf{W}^t = \begin{cases}
\mathbf{B}^\intercal \text{diag}(\mathbf{c}) \mathbf{H}\mathbf{B}, &  (t+1) \bmod q\tau = 0\\
\mathbf{B}^\intercal \text{diag}(\mathbf{c}) \mathbf{B}, & (t+1) \bmod \tau = 0 \\
& \text{ and } (t+1) \bmod q\tau \neq 0\\
\mathbf{I}_{N \times N}, & \text{otherwise,}
\end{cases}
\end{align}
where $\mathbf{B} \in \{0, 1\}^{m \times N}$ is a binary matrix with each element $\mathbf{B}_{i, n}$ denoting if device $n$ belongs to cluster $i$ (i.e., $\mathbf{B}_{i, n} = 1$) or not (i.e., $\mathbf{B}_{i, n} = 0$), $\mathbf{c} = [1/n_1, \ldots, 1/n_m] \in \mathbb{R}^m$, and $\text{diag}(\mathbf{c}) \in \mathbb{R}^{m \times m}$ is a diagonal matrix with the elements of vector $\mathbf{c}$ on the main diagonal. Specifically, for the stage of SGD update (i.e., $(t+1) \bmod \tau \neq 0$), $\mathbf{W}^t$ is the identity matrix because there is no communication between edge devices after SGD update; for the stage of intra-cluster model aggregation (i.e., $(t+1) \bmod \tau = 0$ and $(t+1) \bmod q\tau \neq 0$), $\mathbf{B}^\intercal \text{diag}(\mathbf{c}) \mathbf{B}$ captures the model averaging within each cluster independently after SGD update; and for the stage of inter-cluster model aggregation (i.e., $(t+1) \bmod q\tau = 0$), $\mathbf{B}^\intercal \text{diag}(\mathbf{c}) \mathbf{H}\mathbf{B}$ captures the model aggregation within each cluster followed by a step of gossip averaging across clusters.

To facilitate the convergence analysis, we first introduce the quantities of interests. Multiplying $\mathbf{1}_{N}/N$ on both sides in \eqref{eq_update_rule}, we get 
\begin{equation}
Q(\mathbf{X}^{t+1})\frac{\mathbf{1}_N}{N} =  Q(\mathbf{X}^t) \frac{\mathbf{1}_N}{N} - Q(\eta\mathbf{G}^t) \frac{\mathbf{1}_N}{N},
\end{equation}
where $\mathbf{W}^t$ disappears due to the fact that $\mathbf{1}_N/N$ is a right eigenvector of $\mathbf{B}^\intercal \text{diag}(\mathbf{c}) \mathbf{H}\mathbf{B}$ and $\mathbf{B}^\intercal \text{diag}(\mathbf{c}) \mathbf{B}$ with eigenvalue of 1. 
After rearranging, one can obtain
\begin{equation*}
\mathbf{u}^{t+1} =\mathbf{u}^t-\frac{\eta}{N}\sum_{n=1}^NQ(\mathbf{g}_n^{t}). 
\end{equation*}
\begin{align*}
    \mathbf{g}_n^t = \mathbbm{1}_{n}^t\mathbf{g}_n(\mathbf{x}_n^t),
\end{align*}

where $\mathbbm{1}_{n}^t$ is the Bernoulli random variable satisfying:
\[
    \mathbbm{1}_{n}^t= 
\begin{cases}
    1, & \text{with probability } \rho_n^t\\
    0, & \text{with probability}   (1-\rho_n^t).
\end{cases}
\]
\begin{align*}
    \mathbb{E}[Q(\mathbf{g}_n^t)] = \mathbb{E}[Q(\mathbb{E}_{\mathbbm{1}_{n}^t}[\mathbf{g}_n^t])] = \rho_n^t\mathbb{E}[Q([\mathbf{g}_n(\mathbf{x}_n^{t})])]
\end{align*}
Note that the averaged local model $\mathbf{u}^{t}$ is updated via performing the perturbed SGD contributed by all devices. In the following, we will focus on the convergence of the averaged model $\mathbf{u}^{t}$.

\section{Proof Preliminaries}
For ease of notation, we define the averaged stochastic gradient, the averaged mini-batch gradient, and the averaged gradient of average model as:
\begin{equation*}
    \overline{\mathbf{G}}(\mathbf{X}^{t})=\mathbf{G}_t\frac{\mathbf{1}_{N}}{N}=\frac{1}{N}\sum_{n=1}^N\mathbf{g}_n^t,\quad
    \overline{\nabla F}(\mathbf{X}^{t}) = \mathbb{E}[\overline{\mathbf{G}}^t] = \frac{1}{N}\sum_{n=1}^N\rho_n^t\nabla F_n(\mathbf{x}_{n}^{t}), \quad \nabla F(\mathbf{u}^{t}) =  \frac{1}{N}\sum_{n=1}^N\rho_n^t\nabla F_n(\mathbf{u}^{t}).\\
\end{equation*}

%%%%%%%%%%%%%%%%%%%%%%%%%%%%%%%%%%%%%%%%%%%%%%%%%%%%%%%%%%%%%%%%%%%%%%%%%%%%%%%%%%%%%%%%%%%%%%%%%%%%%%%%%%%%%%%%%%%%%%%%%%%%%%%%%%%%%%%%%%%%%%%%%%%%%%%%%%%%%%%%%%%%%%%%%%%%%%%%%%%%%%%%%%%%%%%%%
%--------------------------------------------------------------------
\section{Useful Lemmas}
%--------------------------------------------------------------------
We use $\mathbf{Z}$ to denote $\mathbf{B}^\intercal \text{diag}(\mathbf{c}) \mathbf{H}\mathbf{B}$. Recall that $\mathbf{A}=\mathbf{B}^\intercal \text{diag}(\mathbf{c}) \mathbf{H}\mathbf{B}$ and $\mathbf{A}=\mathbf{1}_{N}\mathbf{1}_{N}^\intercal/N.$ 

\setcounter{lemma}{2}
\begin{lemma}\label{lemma_eig_zv}
Let the matrices $\mathbf{Z}$ and $\mathbf{A}$ be defined therein. Then we have:
\begin{itemize}
    \item $\mathbf{1}_{n}$ is a right eigenvector of $\mathbf{Z}$ and $\mathbf{A}$ with eigenvalue 1.
    \item $\mathbf{1}_n^\intercal$ is a left eigenvector of $\mathbf{Z}$ and $\mathbf{A}$ with eigenvalue 1.
    \item The eigenvalues of $\mathbf{Z}$ are the same as the eigenvalues of $\mathbf{H}$.
\end{itemize}
\end{lemma}

\begin{lemma}
\label{lemma_comm_A_W}
Let matrices $\mathbf{A}$ and $\mathbf{W}^t$ be defined therein. Then we have:
\begin{align*}
    \mathbf{W}^t\mathbf{A}=\mathbf{A}\mathbf{W}^t=\mathbf{A}.
\end{align*}
\end{lemma}

\begin{lemma}
\label{lemma_eig_Mat}
Let the matrices $\mathbf{Z}$, $\mathbf{A}$, integers $l$ be defined therein. Then we have:
\begin{equation*}
    \|\mathbf{Z}^{l}-\mathbf{A}\|_{\textup{op}}=\zeta^{l},\quad \|\mathbf{A}-\mathbf{A}\|_{\textup{op}}=1,
\end{equation*}
where $\zeta =  \max \{{|\lambda_2(\mathbf{H})|,|\lambda_n(\mathbf{H})|}\}$, and $\lambda_{i}(\cdot)$ denote the $i$-th largest eigenvalue of a matrix.
\end{lemma}

\begin{lemma}
\label{lemma_tr_Fa}
Suppose $\mathbf{C}\in\mathbb{R}^{d\times n}$, $\mathbf{D}\in\mathbb{R}^{n\times n}$ are two matrices, then we have:
\begin{align*}
    \|\mathbf{C}\mathbf{D}\|_{\textup{F}} \leq \|\mathbf{C}\|_{\textup{F}}\|\mathbf{D}\|_{\textup{F}}.
\end{align*}
\end{lemma}

\begin{lemma}
\label{lemma_comm_Z_V}
Let the matrices $\mathbf{Z}$ and $\mathbf{A}$ be defined therein. Then we have:
\begin{align*}
    \mathbf{Z}\mathbf{A}=\mathbf{A}\mathbf{Z}=\mathbf{Z}.
\end{align*}
\end{lemma}

\begin{lemma}
\label{lemma_Fa_op}
Suppose $\mathbf{C}\in\mathbb{R}^{d\times n}$, $\mathbf{D}\in\mathbb{R}^{n\times n}$ are two matrices, then we have:
\begin{align*}
   \|\mathbf{CD}\|_{\textup{F}}  \leq \|\mathbf{C}\|_{\textup{F}}\|\mathbf{D}\|_{\textup{op}}.
\end{align*}
\end{lemma}

Lemmas~\ref{lemma_eig_zv}-\ref{lemma_eig_Mat},~\ref{lemma_tr_Fa}-\ref{lemma_Fa_op} are provided by \cite{sun2023semi} and \cite{castiglia2021multi}, respectively. 

\begin{lemma}
\label{lemma_i_v_op}
Let the matrix $\mathbf{A}$ be defined therein. Then we have:
\begin{align*}
    \|\mathbf{I}-\mathbf{A}\|_{\textup{op}}=1,\quad \|\mathbf{I}-\mathbf{A}\|_{\textup{op}}=1.
\end{align*}
\end{lemma}

\begin{IEEEproof}
According to the definition of the matrix operator norm, we have:
\begin{align*}
    \|\mathbf{I}-\mathbf{A}\|_{\textup{op}} = & \sqrt{\lambda_{\max}(\mathbf{I}-\mathbf{A})^\intercal(\mathbf{I}-\mathbf{A})}\notag\\
     \labelrel={eq_i_v_sqr} & \sqrt{\lambda_{\max}(\mathbf{I}-\mathbf{A})},
\end{align*}
where~\eqref{eq_i_v_sqr} follows from $\mathbf{A}^2=\mathbf{A}$. By using Lemma 5 in \cite{castiglia2021multi}, we have $ \|\mathbf{I}-\mathbf{A}\|_{\textup{op}} =1$. Similarly, we can obtain $\|\mathbf{I}-\mathbf{A}\|_{\textup{op}}=1$. 
\end{IEEEproof} 

%%%%%%%%%%%%%%%%%%%%%%%%%%%%%%%%%%%
\section{Proof of Lemma~\ref{lemma_bound_ave_grad}}\label{append_lemma}
%%%%%%%%%%%%%%%%%%%%%%%%%%%%%%%%%%%

\begin{IEEEproof}
We begin by recalling some preliminary inequalities that will be utilized throughout the proof. Jensen's inequality: $\|\frac{1}{n}\sum_{i=1}^{n}\bm{a}_i\|^2\leq\frac{1}{n}\sum_{i=1}^{n}\|\bm{a}_i\|^2$ for any $\bm{a}_i\in\mathbb{R}^d$. Young's inequality: $\langle\bm{a},\bm{b}\rangle\leq\frac{\gamma\|\bm{a}\|^2}{2}+\frac{\|\bm{b}\|^2}{2\gamma}$ for any $\gamma>0$ and $\bm{a},\bm{b}\in\mathbb{R}^d$. We also have the update rule $
\mathbf{u}^{t+1} =\mathbf{u}^t-\frac{\eta}{N}\sum_{n=1}^NQ(\mathbf{g}_n^{t}).$ For the local compressed gradients, we have: $\mathbb{E}[Q(\mathbf{g}_n^t)] = \mathbb{E}[Q(\mathbb{E}_{\mathbbm{1}_{n}^t}[\mathbf{g}_n^t])] = \rho_n^t\mathbb{E}[Q([\mathbf{g}_n(\mathbf{x}_n^{t})])]$.
According to the Lipschitz Assumption~\ref{ass:smoothness}, we get
\begin{align*}
\mathbb{E}[F(\mathbf{u}^{t+1})
-F(\mathbf{u}^t)] 
\leq\underbrace{\frac{\eta^2L}{2}\mathbb{E}\|\frac{1}{N}\sum_{n=1}^NQ(\mathbf{g}_n^t)\|^2}_{A_1}\underbrace{-\eta\mathbb{E}[\langle\nabla F(\mathbf{u}^{t}),\frac{1}{N}\sum_{n=1}^N\rho_n^tQ(\nabla F_n(\mathbf{x}_n^{t}))\rangle]}_{A_2}.
\end{align*}
For $A_1$, we have
\begin{align*}
    A_1  &= \frac{\eta^2L}{2}\mathbb{E}\|\frac{1}{N}\sum_{n=1}^N [Q(\mathbf{g}_n^t)-\mathbf{g}_n^t+\mathbf{g}_n^t]\|^2\\
    &\leq \eta^2L\mathbb{E}\|\frac{1}{N}\sum_{n=1}^N Q(\mathbf{g}_n^t)-\mathbf{g}_n^t\|^2 + \eta^2L\mathbb{E}\|\frac{1}{N}\sum_{n=1}^N \mathbf{g}_n^t\|^2\\
    &\leq \frac{\eta^2L}{N}\sum_{n=1}^{N}(1-\theta_{n}^{t})\mathbb{E}\|\mathbf{g}_n^t\|^2 + \frac{\eta^2 L}{N}\sum_{n=1}^{N}\mathbb{E}\|\mathbf{g}_n^t\|^2\\
    & \leq \frac{\eta^2 L}{N}\sum_{n=1}^{N}(2-\theta_{n}^t)\mathbb{E}\|\mathbf{g}_n^t\|^2\\
    & = \frac{\eta^2 L}{N}\sum_{n=1}^{n}(2-\theta_{n}^t)\rho_n^t\mathbb{E}\|\mathbf{g}_n(\mathbf{x}_n^{t})\|^2\\
    & \labelrel={eq_a2_exp_var}\frac{\eta^2 L}{N}\sum_{n=1}^{N}(2-\theta_{n}^t)\rho_n^t[\mathbb{E}\|\mathbf{g}_n(\mathbf{x}_n^{t}) - \nabla F_n(\mathbf{x}_n^{t})\|^2 + \frac{\eta^2 L}{N}\sum_{n=1}^{N}(2-\theta_{n}^t)\rho_n^t\|\nabla F_n(\mathbf{x}_n^{t})\|^2]\\
    & = \frac{\eta^2 L}{N}\sum_{n=1}^{N}(2-\theta_{n}^t)\rho_n^t\sigma^2 + \frac{\eta^2 L}{N}\sum_{n=1}^{N}(2-\theta_{n}^t)\rho_n^t\|\nabla F_n(\mathbf{x}_n^{t})\|^2,
\end{align*}
where~\eqref{eq_a2_exp_var} follows from $\mathbb{E}[\|\bm{a}\|^2]=\mathbb{E}[\|\bm{a}-\mathbb{E}(\bm{a})\|^2]+\|\mathbb{E}(\bm{a})\|^2$ with $\bm{a}\in\mathbb{R}^d$.
For $A_2$, we have
\begin{align*}
A_2  &= -\eta\mathbb{E}[\langle\nabla F(\mathbf{u}^{t}),\frac{1}{N}\sum_{n=1}^N\rho_n^t \nabla F_n(\mathbf{x}_n^{t})\rangle]+ \eta\mathbb{E}[\langle\nabla F(\mathbf{u}^{t}),\frac{1}{N}\sum_{n=1}^N\rho_n^t\left[\nabla F_n(\mathbf{x}_n^{t}) -Q(\nabla F_n(\mathbf{x}_n^{t})) \right]\rangle]\\
&\leq -\frac{\eta}{2}\|\nabla F(\mathbf{u}^{t})\|^2 - \frac{\eta}{2N}\sum_{n=1}^N(\rho_n^t)^2\|\nabla F_n(\mathbf{x}_n^{t})\|^2 +\frac{\eta}{2N}\sum_{n=1}^N\|\nabla F(\mathbf{u}^{t})-\rho_n^t \nabla F_n(\mathbf{x}_n^{t})\|^2+\frac{\eta\gamma}{2}\|\nabla F(\mathbf{u}^{t})\|^2\\
& \quad+ \frac{\eta}{2\gamma N}\sum_{n=1}^N(\rho_n^t)^2(1-\theta_n^t)\|\nabla F_n(\mathbf{x}_n^{t})\|^2.
\end{align*}
Combining $A_1$ and $A_2$, we obtain
\begin{align*}
    \mathbb{E}[F(\mathbf{u}^{t+1})-F(\mathbf{u}^t)]  &\leq  \frac{\eta^2 L}{N}\sum_{n=1}^{N}(2-\theta_{n}^t)\rho_n^t\sigma^2+\frac{\eta}{2N}\sum_{n=1}^N\|\nabla F(\mathbf{u}^{t})-\rho_n^t \nabla F_n(\mathbf{x}_n^{t})\|^2-\frac{\eta(1-\gamma)}{2}\|\nabla F(\mathbf{u}^{t})\|^2 \\
    &-\frac{\eta}{2N}\sum_{n=1}^{N}[(\rho_n^t)^2(\frac{\gamma-1+\theta_n^t}{\gamma}) - 2\eta L(2-\theta_n^t)\rho_{n}^{t}]\|\nabla F_n(\mathbf{x}_n^{t})\|^2.
\end{align*}
When $(\rho_n^t)^2-2(\rho_n^t)^2(1-\theta_n^t)\geq 2\eta L(2-\theta_n^t)\rho_{n}^{t}$ and $\gamma<1/2$, we have
\begin{align*}
    &\mathbb{E}[F(\mathbf{u}^{t+1})-F(\mathbf{u}^t)]  \leq \frac{\eta^2 L}{N}\sum_{n=1}^{N}(2-\theta_{n}^t)\rho_n^t\sigma^2  +\frac{\eta}{2N}\sum_{n=1}^N\|\nabla F(\mathbf{u}^{t})-\rho_n^t \nabla F_n(\mathbf{x}_n^{t})\|^2-\frac{\eta}{4}\|\nabla F(\mathbf{u}^{t})\|^2.
\end{align*}
For the term $\sum_{n=1}^N\|\nabla F(\mathbf{u}^{t})-\rho_n^t \nabla F_n(\mathbf{x}_n^{t})\|^2$, we obtain
\begin{align*}
\sum_{n=1}^N\|\nabla F(\mathbf{u}^{t})-\rho_n^t \nabla F_n(\mathbf{x}_n^{t})\|^2&= \sum_{n=1}^N\|\nabla F(\mathbf{u}^{t})-\nabla F_n(\mathbf{x}_n^{t})+(1-\rho_n^t)\nabla F_n(\mathbf{x}_n^{t})\|^2\\
&\leq 2\sum_{n=1}^N\|\nabla F(\mathbf{u}^{t})-\nabla F_n(\mathbf{x}_n^{t})\|^2 + 2\sum_{n=1}^N(1-\rho_n^t)^2\|\nabla F_n(\mathbf{x}_n^{t})\|^2.
\end{align*}
Then, we have
\begin{align*}
     \mathbb{E}[F(\mathbf{u}^{t+1})-F(\mathbf{u}^t)]  &\leq \frac{\eta^2 L}{N}\sum_{n=1}^{N}(2-\theta_{n}^t)\rho_n^t\sigma^2 +\frac{\eta}{N}\sum_{n=1}^N\mathbb{E}\|\nabla F(\mathbf{u}^{t})-\nabla F_n(\mathbf{x}_n^{t})\|^2  -\frac{\eta}{4}\mathbb{E}\|\nabla F(\mathbf{u}^{t})\|^2 \notag\\
    &+ \frac{\eta}{N}\sum_{n=1}^N(1-\rho_n^t)^2\mathbb{E}\|\nabla F_n(\mathbf{x}_n^{t})\|^2\label{in_sum_grad}
\end{align*}
According to Assumption~\ref{ass:smoothness}, we get
\begin{align*}
 \mathbb{E}\|\nabla F(\mathbf{u}^{t})\|^2 
 &\leq  \frac{4[\mathbb{E} F(\mathbf{u}^{t})-\mathbb{E} F(\mathbf{u}^{t+1})]}{\eta}+\frac{4L^2}{N}\sum_{n=1}^{N}\mathbb{E}\|\mathbf{u}^{t}-\mathbf{x}_n^{t}\|^2 + \frac{4\eta L}{N}\sum_{n=1}^{N}(2-\theta_{n}^t)\rho_n^t\sigma^2\\
 &\quad + \frac{12}{N}\sum_{n=1}^N(1-\rho_n^t)^2\mathbb{E}\|\nabla F_n(\mathbf{x}_n^{t})\|^2
\end{align*}
Taking the total expectation and averaging over all iterations, then we arrive at the final result.
\end{IEEEproof}

%---------------------------------
\section{Proof of Lemma~\ref{lemma_bound_disc_error}}\label{append_lemma_bound_disc}
%---------------------------------
\begin{IEEEproof}
From the definition of Frobenius norm and~\eqref{eq:aver_model}, we get:
\begin{align}  
\sum_{n=1}^N\|\mathbf{u}^{t}-\mathbf{x}_{n}^{t}\|^2
    & = \mathbb{E}\|\mathbf{u}^{t}\mathbf{1}_{n}^\intercal-\mathbf{X}^t\mathbf{I}\|_{\textup{F}}^2\notag\\
    & = \mathbb{E}\|\mathbf{X}^t\mathbf{1}_{N}\mathbf{1}_{N}^\intercal-\mathbf{X}^t\mathbf{I}\|_{\textup{F}}^2\notag\\
    & = \mathbb{E}\|\mathbf{X}^t(\mathbf{A}-\mathbf{I})\|_{\textup{F}}^2. 
\end{align}
According to the update rule, we have:
\begin{align}
    \mathbf{X}^t(\mathbf{I}-\mathbf{A}) & =(\mathbf{X}^{t-1}-\eta Q(\mathbf{G}^{t-1}))\mathbf{W}^{t-1}(\mathbf{I}-\mathbf{A})\notag\\
    & \labelrel={eq_intra_stoc} \mathbf{X}^{t-1}(\mathbf{I}-\mathbf{A})\mathbf{W}^{t-1}-\eta Q(\mathbf{G}^{t-1})\mathbf{W}^{t-1}(\mathbf{I}-\mathbf{A})\notag\\
    & = (\mathbf{X}^{t-2}-\eta Q(\mathbf{G}^{t-2}))(\mathbf{I}-\mathbf{A})\mathbf{W}^{t-2}\mathbf{W}^{t-1}-\eta Q(\mathbf{G}^{t-1})\mathbf{W}^{t-1}(\mathbf{I}-\mathbf{A})\notag\\
    & = \mathbf{X}^{t-2}(\mathbf{I}-\mathbf{A})\mathbf{W}^{t-2}\mathbf{W}^{t-1}-\eta 
 Q(\mathbf{G}^{t-2})\mathbf{W}^{t-2}\mathbf{W}^{t-1}(\mathbf{I}-\mathbf{A})-\eta Q(\mathbf{G}^{t-1})\mathbf{W}^{t-1}(\mathbf{I}-\mathbf{A}).\notag
\end{align}
where~\eqref{eq_intra_stoc} follows the special property of doubly stochastic matrix: $\mathbf{A}\mathbf{W}^{t-1} = \mathbf{W}^{t-1}\mathbf{A} = \mathbf{W}^{t-1}$. Then, expanding the expression, we have:
\begin{align}
    \mathbf{X}^t(\mathbf{I}-\mathbf{A}) = \mathbf{X}^0(\mathbf{I}-\mathbf{A})\prod_{u=0}^{t-1}\mathbf{W}^{u}-\eta \sum_{c=1}^{t-1}Q(\mathbf{G}^{c})\prod_{u=c}^{t-1}\mathbf{W}^{u}\left(\mathbf{I}-\mathbf{A}\right).\notag
\end{align}
Here, $u, c$ are the indexes for global iterations. Since all clients were initialized with the same model, $\mathbf{X}^0\prod_{u=1}^{t-1}\mathbf{W}^{u}(\mathbf{I}-\mathbf{A})=0$. Then, the squared norm of intra-cluster residual error can be written as:
\begin{align}\label{eq:L_C}
    \mathbb{E}\|\mathbf{X}^t(\mathbf{I}-\mathbf{A})\|_{\textup{F}}^2 = \eta^2\mathbb{E}\|\sum_{c=1}^{t-1}Q(\mathbf{G}^c)\mathbf{\Phi}^{c,t-1}(\mathbf{I}-\mathbf{A})\|_{\textup{F}}^2,
\end{align} 
where $\mathbf{\Phi}^{c,t-1} := \prod_{u=c}^{t-1}\mathbf{W}^{u}$.
Recall that $t=l q\tau+r\tau+s$, where $l \in [0, \phi - 1]$ is the global round index, $r \in [0, q - 1] $ is the edge round index, and $s \in [0, \tau - 1]$ is the local iteration index. Since $\mathbf{V}^l =\mathbf{V}$ and $\mathbf{VZ}=\mathbf{ZV}=\mathbf{Z}$ by Lemma~\ref{lemma_comm_Z_V}, we have:
\begin{align}
    \mathbf{\Phi}^{c,t-1} = 
        \begin{cases}
            \mathbf{I}, & \ l q\tau+r \tau< c < l q\tau+r \tau+s\\
            \mathbf{V}, & \ l q\tau< c\leq l q\tau+r \tau\\
            \mathbf{Z}, & \ (l -1)q\tau< c \leq l q\tau\\
            \mathbf{Z}^2, & \ (l -2)q\tau< c \leq (l -1)q\tau\\
            \vdots & \\
            \mathbf{Z}^l.  & \ 1\leq c \leq q\tau
        \end{cases}\label{def_Phi_1}
\end{align}
Using the same notations as~\eqref{def_Phi_1} and~\eqref{def_ypqr_1}, we also let
\begin{equation}\label{def_ypqr_1}
    \mathbf{Y}_{\alpha}  =\sum_{c=\alpha q\tau+1}^{(\alpha+1)q\tau}Q(\mathbf{G}^c), \mathbf{Y}_{l,r}=\sum_{c=l q\tau+1}^{l q\tau+r \tau}Q(\mathbf{G}^c), \mathbf{Y}_{l,r,s}=\sum_{c=l q\tau+r\tau+1}^{l q\tau+r \tau+s-1}Q(\mathbf{G}^c).
\end{equation}

Thus, we obtain:
\begin{align}
    \sum_{c=1}^{q\tau}Q(\mathbf{G}^c)\Phi^{c,t-1}(\mathbf{I}-\mathbf{A}) & =\mathbf{Y}_0\mathbf{Z}^l(\mathbf{I} -\mathbf{A}),\notag\\
    \sum_{c=q\tau+1}^{2q\tau}Q(\mathbf{G}^c)\Phi^{c,t-1}(\mathbf{I}-\mathbf{A}) & =\mathbf{Y}_1\mathbf{Z}^{l -1}(\mathbf{I}-\mathbf{A}),\notag\\
    & \ldots, \notag\\
    \sum_{c=(l -1)q\tau+1}^{l q\tau}Q(\mathbf{G}^c)\Phi^{c,t-1}(\mathbf{I}-\mathbf{A}) & =\mathbf{Y}_{l -1}\mathbf{Z}(\mathbf{I}-\mathbf{A}),\notag\\
    \sum_{c=l q\tau+1}^{l q\tau+r \tau+s-1}Q(\mathbf{G}^c)\Phi^{c,t-1}(\mathbf{I}-\mathbf{A}) & = \mathbf{Y}_{l ,r}\mathbf{V}(\mathbf{I}-\mathbf{A}) + \mathbf{Y}_{l ,r, s}\mathbf{I}(\mathbf{I}-\mathbf{A}).\notag
\end{align}
By summing them all together and Lemmas~\ref{lemma_comm_A_W},~\ref{lemma_comm_Z_V}, we get:
\begin{align}\label{EQ:SUM_L_C}
    \sum_{c=1}^{t-1}Q(\mathbf{G}_{c})\mathbf{\Phi}_{c,t-1}(\mathbf{I}-\mathbf{A})=\sum_{\alpha =0}^{l -1}\mathbf{Y}_{\alpha }(\mathbf{Z}^{l -\alpha }-\mathbf{A})+\mathbf{Y}_{l,r}(\mathbf{V}-\mathbf{A})+\mathbf{Y}_{l,r,s}(\mathbf{I}-\mathbf{A}).
\end{align}

Plugging~\eqref{EQ:SUM_L_C} into~\eqref{eq:L_C}, we obtain:
\begin{align}
    \mathbb{E}&\|\mathbf{X}_t(\mathbf{I}-\mathbf{A})\|_{\textup{F}}^2 =  \eta^2\mathbb{E}\|\sum_{\alpha =0}^{l -1}\mathbf{Y}_{\alpha }(\mathbf{Z}^{l -\alpha }-\mathbf{A})+\mathbf{Y}_{l,r}(\mathbf{V}-\mathbf{A})+\mathbf{Y}_{l,r,s}(\mathbf{I}-\mathbf{A})\|_{\textup{F}}^2\notag\\
    =& \eta^2\left(\sum_{\alpha =0}^{l -1}\mathbb{E}\|\mathbf{Y}_{\alpha }(\mathbf{Z}^{l -\alpha }-\mathbf{A})\|^2 + \mathbb{E}\|\mathbf{Y}_{l,r}(\mathbf{V}-\mathbf{A})\|^2 + \mathbb{E}\|\mathbf{Y}_{l,r,s}(\mathbf{I}-\mathbf{A})\|_{\textup{F}}^2\right)\notag\\
     & + \underbrace{\eta^2\sum_{\alpha=0}^{l -1}\sum_{\alpha^{\prime}=0,\alpha^{\prime}\neq \alpha}^{l -1}\mathbb{E}\underbrace{\left<\mathbf{Y}_{\alpha}(\mathbf{Z}^{l -\alpha}-\mathbf{A}), \mathbf{Y}_{\alpha^{\prime}}(\mathbf{Z}^{l-{\alpha^{\prime}}}-\mathbf{A}) \right>}_{TR}}_{TR_0}+ \underbrace{2\eta^2\sum_{\alpha=0}^{l -1}\mathbb{E}\left<\mathbf{Y}_{l,r}(\mathbf{V}-\mathbf{A}), \mathbf{Y}_\alpha(\mathbf{Z}^{l -\alpha}-\mathbf{A})  \right>}_{TR_1}\notag\\
    & +\underbrace{2\eta^2\sum_{\alpha=0}^{l -1}\mathbb{E}\left<\mathbf{Y}_{l,r,s}(\mathbf{I}-\mathbf{A}), \mathbf{Y}_\alpha(\mathbf{Z}^{l -\alpha}-\mathbf{A})  \right>}_{TR_2} + \underbrace{2\eta^2\mathbb{E}\left<\mathbf{Y}_{l,r}(\mathbf{V}-\mathbf{A}), \mathbf{Y}_{l,r,s}(\mathbf{I}-\mathbf{A})  \right>}_{TR_3}\label{eq_t3_tr}
\end{align}
$\emph{TR}$ can be bounded as:
\begin{align}
    \emph{TR} \labelrel\leq{in_T3_TR} & \mathbb{E}\|\mathbf{Y}_\alpha(\mathbf{Z}^{l -\alpha}-\mathbf{A})\|_{\textup{F}} \mathbb{E}\|\mathbf{Y}_{\alpha^{\prime}}(\mathbf{Z}^{l -\alpha^{\prime}}-\mathbf{A})\|_{\textup{F}}\notag\\
    \labelrel\leq{in_T4_FA_OP_1} & \mathbb{E}\|\mathbf{Y}_\alpha\|_{\textup{F}}\|\mathbf{Z}^{l -\alpha}-\mathbf{A}\|_{\textup{op}} \mathbb{E}\|\mathbf{Y}_{\alpha^{\prime}}\|_{\textup{F}}\|\mathbf{Z}^{l -{\alpha^{\prime}}}-\mathbf{A}\|_{\textup{op}}\notag\\
    \labelrel\leq{in_T4_OP_EIG_1} & \zeta^{(2l-\alpha-{\alpha^{\prime}})}\mathbb{E}\|\mathbf{Y}_\alpha\|_{\textup{F}}\mathbb{E}\|\mathbf{Y}_{\alpha^{\prime}}\|_{\textup{F}}\notag\\
    \leq & \frac{1}{2}\zeta^{(2l-\alpha-{\alpha^{\prime}})}[\mathbb{E}\|\mathbf{Y}_{\alpha}\|_{\textup{F}}^2+\mathbb{E}\|\mathbf{Y}_{\alpha^{\prime}}\|_{\textup{F}}^2],\notag
\end{align}
where~\eqref{in_T3_TR} follows from Cauchy-Schwarz inequality,~\eqref{in_T4_FA_OP_1} follows from Lemma~\ref{lemma_Fa_op} and~\eqref{in_T4_OP_EIG_1} follows from Lemma~\ref{lemma_eig_Mat}.
Using the same techniques, we get:
\begin{align*}
    \emph{TR}_1 &\leq \eta^2\sum_{\alpha=0}^{l -1}\zeta^{(l-{\alpha})}[\mathbb{E}\|\mathbf{Y}_{l,r}\|_{\textup{F}}^2+\mathbb{E}\|\mathbf{Y}_{\alpha}\|_{\textup{F}}^2],\\
    \emph{TR}_2 &\leq \eta^2\sum_{\alpha=0}^{l -1}\zeta^{(l-{\alpha})}[\mathbb{E}\|\mathbf{Y}_{l,r,s}\|_{\textup{F}}^2+\mathbb{E}\|\mathbf{Y}_{\alpha}\|_{\textup{F}}^2],\\
    \emph{TR}_3 &\leq \eta^2[\mathbb{E}\|\mathbf{Y}_{l,r,s}\|_{\textup{F}}^2+\mathbb{E}\|\mathbf{Y}_{l,r}\|_{\textup{F}}^2].
\end{align*}
By summing the above inequalities of $\emph{TR}_0, \emph{TR}_1, \emph{TR}_2$ and $\emph{TR}_3$, we have:
\begin{align}
    \emph{TR}_0+\emph{TR}_1+\emph{TR}_2+\emph{TR}_3 & \leq  \eta^2\sum_{\alpha=0}^{l -1}\sum_{{\alpha^{\prime}}=0,\alpha^{\prime}\neq \alpha}^{l-1}\zeta^{(2l-\alpha-{\alpha^{\prime}})}[\mathbb{E}\|\mathbf{Y}_\alpha\|_{\textup{F}}^2+\mathbb{E}\|\mathbf{Y}_{\alpha^{\prime}}\|_{\textup{F}}^2] + \eta^2\sum_{\alpha=0}^{l -1}\zeta^{(l-\alpha)}[\mathbb{E}\|\mathbf{Y}_{l,r}\|_{\textup{F}}^2+\mathbb{E}\|\mathbf{Y}_\alpha\|_{\textup{F}}^2]\notag\\
    & \quad+ \eta^2\sum_{\alpha=0}^{l -1}\zeta^{(l-\alpha)}\left[\mathbb{E}\|\mathbf{Y}_{l,r,s}\|_{\textup{F}}^2+\mathbb{E}\|\mathbf{Y}_\alpha\|_{\textup{F}}^2\right]+ \eta^2\mathbb{E}\|\mathbf{Y}_{l,r,s}\|_{\textup{F}}^2 + \eta^2\mathbb{E}\|\mathbf{Y}_{l,r}\|_{\textup{F}}^2.\notag\\
    & = \eta^2\sum_{\alpha=0}^{l -1}\zeta^{l-\alpha}\mathbb{E}\|\mathbf{Y}_\alpha\|_{\textup{F}}^2\sum_{{\alpha^{\prime}}=0,\alpha^{\prime}\neq\alpha}^{l-1}\zeta^{(l-{\alpha^{\prime}})} + 2\eta^2\sum_{\alpha=0}^{l-1}\zeta^{l-\alpha}\mathbb{E}\|\mathbf{Y}_\alpha\|_{\textup{F}}^2\notag\\
    &\quad + \eta^2\sum_{\alpha=0}^{l}\zeta^{(l-\alpha)}\mathbb{E}\|\mathbf{Y}_{l,r}\|_{\textup{F}}^2 + \eta^2\sum_{\alpha=0}^{l}\zeta^{(l-\alpha)}\mathbb{E}\|\mathbf{Y}_{l,r,s}\|_{\textup{F}}^2\notag\\
    & \labelrel\leq{in_t3_zeta} \eta^2\sum_{\alpha=0}^{l -1}\zeta^{l-\alpha}\mathbb{E}\|\mathbf{Y}_\alpha\|_{\textup{F}}^2\frac{\zeta}{1-\zeta} + 2\eta^2\sum_{\alpha=0}^{l-1}\zeta^{l-\alpha}\mathbb{E}\|\mathbf{Y}_\alpha\|_{\textup{F}}^2\notag\\
    &\quad + \eta^2\mathbb{E}\|\mathbf{Y}_{l,r}\|_{\textup{F}}^2\frac{1}{1-\zeta} + \eta^2\mathbb{E}\|\mathbf{Y}_{l,r,s}\|_{\textup{F}}^2\frac{1}{1-\zeta}\label{in_sum_t0_3}
\end{align}
where~\eqref{in_t3_zeta} uses the following fact:
\begin{align*}
        \sum_{\alpha=0}^{l} \zeta^{l -\alpha} \leq \sum_{\alpha=-\infty}^{l-1}\zeta^{l -\alpha} \leq \frac{1}{1-\zeta},\quad \sum_{\alpha=0}^{l-1} \zeta^{l -\alpha} \leq \sum_{\alpha=-\infty}^{l-1}\zeta^{l -\alpha} \leq \frac{\zeta}{1-\zeta}.
\end{align*}
Plugging~\eqref{in_sum_t0_3} back into~\eqref{eq_t3_tr}, we have:
\begin{align}
    \mathbb{E}\|\mathbf{X}_t(\mathbf{I}-\mathbf{A})\|_{\textup{F}}^2 & \leq \eta^2\sum_{\alpha =0}^{l -1}\mathbb{E}\|\mathbf{Y}_{\alpha }\|_{\textup{F}}^2\zeta^{2(l-\alpha)} + \eta^2\mathbb{E}\|\mathbf{Y}_{l,r}\|_{\textup{F}}^2 + \frac{\eta^2}{n}\mathbb{E}\|\mathbf{Y}_{l,r,s}\|_{\textup{F}}^2\notag\\
    & \quad +\eta^2\sum_{\alpha=0}^{l -1}\zeta^{l-\alpha}\mathbb{E}\|\mathbf{Y}_\alpha\|_{\textup{F}}^2\frac{\zeta}{1-\zeta} + 2\eta^2\sum_{\alpha=0}^{l-1}\zeta^{l-\alpha}\mathbb{E}\|\mathbf{Y}_\alpha\|_{\textup{F}}^2\notag\\
    &\quad + \eta^2\mathbb{E}\|\mathbf{Y}_{l,r}\|_{\textup{F}}^2\frac{1}{1-\zeta} + \eta^2\mathbb{E}\|\mathbf{Y}_{l,r,s}\|_{\textup{F}}^2\frac{1}{1-\zeta}\notag\\
    & \leq \eta^2\sum_{\alpha =0}^{l -1}\left(\zeta^{2(l-\alpha)}+2\zeta^{l-\alpha}+\frac{\zeta^{l-\alpha+1}}{1-\zeta} \right)\mathbb{E}\|\mathbf{Y}_{\alpha }\|_{\textup{F}}^2 + \eta^2\left(\frac{2-\zeta}{1-\zeta}\right)\mathbb{E}\|\mathbf{Y}_{l,r}\|_{\textup{F}}^2 + \eta^2\left(\frac{2-\zeta}{1-\zeta}\right)\mathbb{E}\|\mathbf{Y}_{l,r,s}\|_{\textup{F}}^2\label{bound_T3_Y}
\end{align}
Taking a closer look at $\mathbb{E}\|\mathbf{Y}_{\alpha }\|_{\textup{F}}^2$ for $0\leq\alpha<l-1$:
\begin{align}
    \mathbb{E}\|\mathbf{Y}_{\alpha }\|_{\textup{F}}^2 & = \mathbb{E}\|\sum_{c=\alpha q\tau +1}^{(\alpha+1)q\tau}Q(\mathbf{G}_{c})\|_{\textup{F}}^2\notag\\
    & = \sum_{n=1}^{N}\mathbb{E}\|\sum_{c=\alpha q\tau +1}^{(\alpha+1)q\tau}Q(\mathbf{g}_{k}^c)\|^2\notag\\
    & \leq \sum_{n=1}^{N}q\tau\sum_{c=\alpha q\tau +1}^{(\alpha+1)q\tau}\mathbb{E}\|Q(\mathbf{g}_{n}^c)\|^2\label{bound_Y_alpha}
\end{align}
Using similar techniques, for the global round $l$, we obtain:
\begin{align}
    \mathbb{E}\|\mathbf{Y}_{l,r}\|_{\textup{F}}^2 & \leq \sum_{n=1}^{N}r\tau\sum_{c=l q\tau+1}^{l q\tau+r\tau}\mathbb{E}\|Q(\mathbf{g}_{n}^c)\|^2,\label{bound_Y_lr}\\
    \mathbb{E}\|\mathbf{Y}_{l,r,s}\|_{\textup{F}}^2 & \leq \sum_{n=1}^{N}(s-1)\sum_{c=l q\tau+r\tau+1}^{l q\tau+r \tau+s-1}\mathbb{E}\|Q(\mathbf{g}_{n}^c)\|^2.\label{bound_Y_lrs}
\end{align}
Plugging~\eqref{bound_Y_alpha},~\eqref{bound_Y_lr} and~\eqref{bound_Y_lrs} back into~\eqref{bound_T3_Y}, we get:
\begin{align}
    \mathbb{E}\|\mathbf{X}_t(\mathbf{I}-\mathbf{A})\|_{\textup{F}}^2 \leq & \eta^2q\tau\sum_{\alpha =0}^{l -1}\left(\zeta^{2(l-\alpha)}+2\zeta^{l-\alpha}+\frac{\zeta^{l-\alpha+1}}{1-\zeta} \right)\sum_{n=1}^{N}\sum_{c=\alpha q\tau +1}^{(\alpha+1)q\tau}\mathbb{E}\|Q(\mathbf{g}_{n}^c)\|^2 + \eta^2r\tau\left(\frac{2-\zeta}{1-\zeta}\right)\sum_{n=1}^{N}\sum_{c=l q\tau+1}^{l q\tau+r\tau}\mathbb{E}\|Q(\mathbf{g}_{n}^c)\|^2\notag\\
    & \quad + \eta^2(s-1)\left(\frac{2-\zeta}{1-\zeta}\right)\sum_{c=l q\tau+r\tau+1}^{l q\tau+r \tau+s-1}\mathbb{E}\|Q(\mathbf{g}_{n}^c)\|^2\notag
\end{align}
Our goal is to calculate $\sum_{t=0}^{T-1}\mathbb{E}\|\mathbf{X}_t(\mathbf{I}-\mathbf{A})\|_{\textup{F}}^2$, which can be decomposed into $\sum_{l=0}^{p-1}\sum_{r=0}^{q-1}\sum_{s=0}^{\tau-1}\mathbb{E}\|\mathbf{X}_t(\mathbf{I}-\mathbf{A})\|_{\textup{F}}^2$. We sum over all iterates in $l $-th global round period, for $r=0, \ldots ,q-1$ and $s=0, \ldots ,\tau-1$:
\begin{align}
    \sum_{r=0}^{q-1}\sum_{s=0}^{\tau-1}\mathbb{E}\|\mathbf{X}_t(\mathbf{I}-\mathbf{A})\|_{\textup{F}}^2 &\leq \eta^2q^2\tau^2\sum_{\alpha =0}^{l -1}\left(\zeta^{2(l-\alpha)}+2\zeta^{l-\alpha}+\frac{\zeta^{l-\alpha+1}}{1-\zeta} \right)\sum_{n=1}^{N}\sum_{c=\alpha q\tau +1}^{(\alpha+1)q\tau}\mathbb{E}\|Q(\mathbf{g}_{n}^c)\|^2\notag\\
    &\quad + \eta^2\frac{q(q-1)}{2}\tau^2\left(\frac{2-\zeta}{1-\zeta}\right)\sum_{n=1}^{N}\sum_{c=l q\tau+1}^{(l+1)q\tau}\mathbb{E}\|Q(\mathbf{g}_{n}^c)\|^2\notag\\
    & \quad + \eta^2q\frac{\tau(\tau-1)}{2}\left(\frac{2-\zeta}{1-\zeta}\right)\sum_{n=1}^{N}\sum_{c=l q\tau+1}^{(l+1) q\tau}\mathbb{E}\|Q(\mathbf{g}_{n}^c)\|^2.\label{bound_T3_ZETA_1}
\end{align}
Here, for $\alpha \in [0,\dots,l]$, we denote:
\begin{equation}\label{eq_Gamma}
    \Gamma_\alpha  =  \zeta^{2(l -\alpha )}+2\zeta^{l -\alpha }+\frac{\zeta^{(l -\alpha +1)}}{1-\zeta}.
\end{equation}
We can find that $\Gamma_l  = \frac{3-\zeta}{1-\zeta}>\frac{2-\zeta}{1-\zeta}$. Thus, we can further bound~\eqref{bound_T3_ZETA_1} as
\begin{align}
    \sum_{r=0}^{q-1}\sum_{s=0}^{\tau-1}\mathbb{E}\|\mathbf{X}_t(\mathbf{I}-\mathbf{A})\|_{\textup{F}}^2 \leq \eta^2q^2\tau^2\sum_{\alpha =0}^{l }\Gamma_\alpha \sum_{n=1}^{N}\sum_{c=l q\tau+1}^{(l+1)q\tau}\mathbb{E}\|Q(\mathbf{g}_{n}^c)\|^2.\notag
\end{align}
Then, summing over all global rounds for $l =0,\ldots,p-1$, we have:
\begin{align}
    \sum_{l =0}^{p-1}\sum_{r=0}^{q-1}\sum_{s=0}^{\tau-1}\mathbb{E}\|\mathbf{X}_t(\mathbf{I}-\mathbf{A})\|_{\textup{F}}^2\leq & \eta^2 q^2\tau^2\sum_{l=0}^{p-1}\sum_{\alpha =0}^{l }\Gamma_\alpha \sum_{n=1}^{N}\sum_{c=\alpha  q\tau+1}^{(\alpha +1)q\tau}\mathbb{E}\|Q(\mathbf{g}_{n}^c)\|^2\notag\\
    = & \eta^2 q^2\tau^2\sum_{\alpha =0}^{p-1}\sum_{l=\alpha }^{p-1 }\Gamma_{l}\sum_{n=1}^{N}\sum_{c=\alpha  q\tau+1}^{(\alpha +1)q\tau}\mathbb{E}\|Q(\mathbf{g}_{n}^c)\|^2.    \label{eq_delta_f}
\end{align}
Expanding the summation in~\eqref{eq_delta_f}, we get:
\begin{align}
    \sum_{l =\alpha }^{p-1}\left(  \zeta^{2(l -\alpha )}+2\zeta^{l -\alpha }+\frac{\zeta^{l -\alpha +1}}{1-\zeta}\right) &\leq \sum_{l =\alpha }^{\infty}\left(  \zeta^{2(l -\alpha )}+2\zeta^{l -\alpha }+\frac{\zeta^{l -\alpha +1}}{1-\zeta}\right)\notag\\
    & \leq \frac{1}{1-\zeta}+\frac{2}{1-\zeta}+\frac{\zeta}{(1-\zeta)^{2}}.\notag
\end{align}
Let $\Omega_{1}=\frac{1}{1-\zeta^{2}}+\frac{2}{1-\zeta}+\frac{\zeta}{(1-\zeta)^2}$. Then we have:
\begin{align*}
   \sum_{t=0}^{T-1}\mathbb{E}\|\mathbf{X}_t(\mathbf{I}-\mathbf{A})\|_{\textup{F}}^2\leq \eta^2q^2\tau^2 \Omega_{1} \sum_{t=0}^{T-1}\sum_{n=1}^{N}\mathbb{E}\|Q(\mathbf{g}_{n}^t)\|^2.
\end{align*}
For the term $\mathbb{E}\|Q(\mathbf{g}_{n}^t)\|^2$, we have
\begin{align*}
\sum_{n=1}^{N}\mathbb{E}\|Q(\mathbf{g}_{n}^t)\|^2  & = \sum_{n=1}^{N}\mathbb{E}\|Q(\mathbf{g}_n^t)-\mathbf{g}_n^t+\mathbf{g}_n^t\|^2\\
    &\leq 2\sum_{n=1}^{N}\mathbb{E}\|Q(\mathbf{g}_n^t)-\mathbf{g}_n^t\|^2 + 2\sum_{n=1}^{N}\mathbb{E}\|\mathbf{g}_n^t\|^2\\
    &\labelrel\leq{in_lemma_1} 2\sum_{n=1}^{N}(1-\theta_{n}^{t})\mathbb{E}\|\mathbf{g}_n^t\|^2 + 2\sum_{n=1}^{N}\mathbb{E}\|\mathbf{g}_n^t\|^2\\
    & = 2\sum_{n=1}^{N}(2-\theta_{n}^t)\mathbb{E}\|\mathbf{g}_n^t\|^2\\
    & = 2\sum_{n=1}^{N}(2-\theta_{n}^t)\rho_n^t\mathbb{E}\|\mathbf{g}_n(\mathbf{x}_n^{t})\|^2\\
    & \leq 4\sum_{n=1}^{N}(2-\theta_{n}^t)\rho_n^t\big[\mathbb{E}\|\mathbf{g}_n(\mathbf{x}_n^{t}) - \nabla F_n(\mathbf{x}_n^{t})\|^2 \\
    &\quad + 4\sum_{n=1}^{N}(2-\theta_{n}^t)\rho_k^t\|\nabla F_n(\mathbf{x}_n^{t})\|^2]\\
    & \labelrel\leq{in_lemma_2} 4\sum_{n=1}^{N}(2-\theta_{n}^t)\rho_n^t\sigma^2 + 4\sum_{n=1}^{N}(2-\theta_{n}^t)\rho_n^t\|\nabla F_n(\mathbf{x}_n^{t})\|^2,
\end{align*}
where~\eqref{in_lemma_1} holds due to the property of $\topk_k$
compression operator,~\eqref{in_lemma_2} follows from Assumption~\ref{ass:gradient}.
In final, we have the following 
\begin{equation}
    \sum_{t=0}^{T-1}\sum_{n=1}^N\mathbb{E}\|\mathbf{u}^{t}-\mathbf{x}_{n}^{t}\|^2 \leq 4\eta^2q^2\tau^2 \Omega_{1} \sum_{t=0}^{T-1}\sum_{n=1}^{N}\big[(2-\theta_{n}^t)\rho_n^t\sigma^2 + (2-\theta_{n}^t)\rho_n^t\|\nabla F_n(\mathbf{x}_n^{t})\|^2\big].
\end{equation}
\end{IEEEproof}
Lemma~\ref{lemma_bound_disc_error} gives the upper bound of the discrepancy error term. Combining Lemmas~\ref{lemma_bound_ave_grad},~\ref{lemma_bound_disc_error} and choosing a proper learning rate, we arrive at the following final convergence bound.
%----------
\section{Proof of Theorem~\ref{th:convergence}}\label{append:theorem}
%-----------
\begin{IEEEproof}
Plugging Lemma~\ref{lemma_bound_disc_error} into Lemma~\ref{lemma_bound_ave_grad}, and in light of Assumption~\ref{ass:bounded_grad}, we get
\begin{align*}
 \frac{1}{T}\sum_{t=0}^{T-1}\mathbb{E}\|\nabla F(\mathbf{u}^{t})\|^2 &\leq\frac{4( F(\mathbf{x}^1)- F_{\inf})}{\eta T} + \frac{4\eta L\sigma^2}{NT}\sum_{t=0}^{T-1}\sum_{n=1}^{N}\left[(2-\theta_n^t)\rho_n^t + 4\eta L q^2\tau^2\Omega_{1}(2-\theta_{n}^t)\rho_n^t)\right]\notag\\
 &+\frac{4G^2}{NT}\sum_{t=0}^{T-1}\sum_{n=1}^{N}\left[3(1-\rho_n^t)^2 + 4L^2\eta^2q^2\tau^2\Omega_{1}(2-\theta_{n}^t)\rho_n^t\right].
\end{align*}
When $\eta \leq \frac{1}{4Lq^2\tau^2\Omega_1}$, we have $2L^2\eta^2q^2\tau^2\Omega_{1} \leq 1$, $4L\eta q^2\tau^2\Omega_{1} \leq 1$. After turning the iteration index $t$ by $\phi$, $q$, and $\tau$, we arrive at the final result.
\end{IEEEproof}

\end{document}